\documentclass[manuscript,screen]{acmart} 

\usepackage{graphicx}
\usepackage[inline]{enumitem}
\usepackage{wrapfig}
\usepackage{makecell}
\usepackage{pifont}
\usepackage{amsthm}
\usepackage{amsmath}
\usepackage{mathtools}
\usepackage{stmaryrd} 
\usepackage{amsfonts}
\usepackage{flushend}
\usepackage{balance}
\usepackage{hyperref}
\usepackage{multicol}
\usepackage{multirow}
\usepackage{setspace} 
\usepackage{xspace}
\usepackage{listings}
\usepackage{ifthen}
\usepackage{verbatim}
\usepackage{float} 
\usepackage{subcaption}
\usepackage{microtype}
\usepackage{mathptmx} 
\usepackage{textcomp}
\usepackage{booktabs}
\usepackage{longtable}
\usepackage{blindtext}
\usepackage{fancyhdr}
\usepackage[mathletters]{ucs}
\usepackage[utf8]{inputenc}
\usepackage{wasysym}	
\usepackage{tikz}
\usepackage{esvect}
\usepackage{csquotes}
\usepackage{array}
\usepackage{xcolor, colortbl}
\usepackage{tablefootnote}
\usepackage{supertabular}
\usepackage[]{mdframed}

\usepackage{algorithm}
\usepackage{algorithmicx}
\usepackage{algpseudocode}

 \newcolumntype{L}{>{$}l<{$}}
 	\newcolumntype{C}{>{$}c<{$}}

\newcommand{\etal}{\hbox{\emph{et al.}}\xspace}

\newcommand{\ie}{\hbox{\emph{i.e.}}\xspace}

\theoremstyle{remark}

\theoremstyle{definition}

\DeclareUnicodeCharacter{183}{\cdot}						
\DeclareUnicodeCharacter{931}{\ensuremath\Sigma}			
\DeclareUnicodeCharacter{9001}{\ensuremath\langle}			
\DeclareUnicodeCharacter{9002}{\ensuremath\rangle}			
\DeclareUnicodeCharacter{9608}{\ensuremath\blacksquare}		
\DeclareUnicodeCharacter{1013}{\in}							
\DeclareUnicodeCharacter{8213}{---}							

\usepackage{framed}
\newcounter{RQCounter}

\DeclareTextFontCommand{\findings}{\normalfont\itshape\bfseries}

\newlength{\emstr}
\newboolean{showcomments}
\setboolean{showcomments}{true} 
\ifthenelse{\boolean{showcomments}}{
  \newcommand{\nbc}[3]{
    {\colorbox{#3}{\bfseries\sffamily\scriptsize\textcolor{white}{#1}}}%
    {\textcolor{#3}{\sf\small
    \textit{#2}
    }}}
  \newcommand{\todo}[1]{\nbc{TODO}{#1}{blue}\xspace}
  \newcommand{\mar}[1]{\nbc{MAR}{#1}{cyan}\xspace}
  \newcommand{\fs}[1]{\nbc{FEDERICA}{#1}{violet}\xspace}
  \newcommand{\carlos}[1]{\nbc{CARLOS}{#1}{olive}\xspace}
}{
  \newcommand{\nbc}[3]{}
  \newcommand{\todo}[1]{}
  \newcommand{\mar}[1]{}
  \newcommand{\fs}[1]{}
  \newcommand{\carlos}[1]{}
}

\begin{document}

\title{The Quest for Content: A Survey of Search-Based Procedural Content Generation for Video Games}

\author{Mar Zamorano}
 \affiliation{
    \institution{University College of London}
    \country{UK}}
\affiliation{
    \institution{Universidad San Jorge}
    \country{Spain}}
\email{maria.lopez.20@ucl.ac.uk}
\email{mzamorano@usj.es}

\author{Carlos Cetina}
\affiliation{
    \institution{University College of London}
    \country{UK}}
\affiliation{
    \institution{Universidad San Jorge}
    \country{Spain}}
\email{ccetina@usj.es}

\author{Federica Sarro}
\affiliation{
    \institution{University College of London}
    \country{UK}}
\email{f.sarro@ucl.ac.uk}

\renewcommand{\shortauthors}{Zamorano et al.}




\begin{abstract}
Video games demand is constantly increasing, which requires the costly production of large amounts of content. Towards this challenge, researchers have developed Search-Based Procedural Content Generation (SBPCG), that is, the (semi-)automated creation of content through search algorithms. We survey the current state of SBPCG, reporting work appeared  in the field between 2011-2022 and identifying open research challenges. The results lead to recommendations for practitioners and to the identification of several potential future research avenues for SBPCG.

\end{abstract}




\maketitle
 
\section{Introduction}
\label{sec:introduction}

Video games are complex software products where art and coding are combined during the development process to conform the final product. The industrial scene of video games development has seen a rapid expansion in the last decades, with video games becoming an economic motor and a worldwide phenomenon that has captured the general interest of all sectors of the population globally~\cite{engelstatter2022video}.

The industry of video games development can be split into three main different branches: AAA games, indie games, and serious games.
The main players in the field develop AAA video games, which are produced and distributed by mid-sized or major publishers~\cite{mathews2016modern}. These games are typically more complex than other types of games, and have high development and marketing budgets, while being mainly constrained by time factors in the form of deadlines and firm release dates.  On the other side of the spectrum, it is possible to find the so-called indie games, created by individuals or smaller video game studios that lack the financial and technical support of a large game publisher. While indie games benefit from unbridled creativity, they often struggle with funding and issues typically associated to small development teams such as having a constrained extension or undesired functionality due to limited testing. 
Serious games are designed not only to entertain, but to also instruct, inform, or educate the players. We can find examples of serious games in Non-Governmental Organizations that develop games to raise awareness about societal issues, medical research associations that do so to simulate complex operations, or top-valued tech companies that use gamification to instruct new recruits in their processes. These games are usually restricted to the scope of their respective domain, and can vary wildly in their goals, complexity, and development times.

Regardless of the differences in their development goals and processes, all these games share a common issue in the form of their \textit{need for content}. 
Of all games, AAA games are usually the ones that require the most content both in amount and variety, often resulting in a long, manual, repetitive, and very costly content creation effort that is carried out by large amounts of game developers. On the other hand, in indie games, content creation is a delicate process constrained to the capabilities of the team. While being too ambitious with game content can cause delays in a project or even result in failure to publish a game, being too lax with the content can render a game unsuccessful. For small game studios, as it may happen with any start-up company, both scenarios can be deadly due to the limited available funding.  Serious games are in the same spot as indie games with regard to their limits in content creation, but often for different reasons. In the case of serious games, their very nature narrows their scope and their target players, and restricts the creation of content to that which is strictly necessary within the domain of the application. While this may be seen as beneficial, it is more often than not a severe hindrance in development, since the specificity of content often calls for unique solutions involving usually unavailable domain experts.

To tackle these development challenges, researchers have focused their efforts towards Procedural Content Generation~\cite{shaker2013automatic}, a field of work that has gained traction in the last decades. 
\textbf{Procedural Content Generation} (\textbf{PCG})~\cite{Hendrikx2013} is defined as the automated production of different components of video game. Through PCG, video game software engineers generate components and aspects of video games in an automated or semi-automated fashion, which is a more competitive approach than doing so through the traditional manual development process. 

PCG can be carried out in two different manners, corresponding to two different stages of the lifetime of a game. Hence, offline PCG refers to content generated before the release of a game (at design time), and online PCG refers to content generated on the fly while the game is being played (at run-time). 

PCG is a large field spanning many algorithms~\cite{yannakakis2018artificial}, which can be grouped in three main categories according to the survey of PCG techniques by Barriga et al.~\cite{Barriga2019}: Traditional methods (TM) that generate content under a procedure without evaluation; Machine Learning (ML) methods that train models to generate new content; and Search-Based (SB) methods that generate content through a search on a predefined space guided by a meta-heuristic using one or more objective functions.

This work surveys the current state of \textbf{Search-Based PCG} (\textbf{SBPCG)} and provide insights for future directions in this field. The ultimate aims is to encourage further applications of Search-Based Software Engineering to Game Software Engineering. as we may have just scratched the surface of its application in this area. In the following subsection we further detail the scope of our survey.

\subsection{Scope}
\label{subsec:scope}
Traditional methods include diverse methods, such as pseudo-random number generators, cellular automata, generative grammars, fractals, or noise. Developers must design an approach for a specific type of content which produces useful elements for that type of content without an evaluation or a learning process. Thus, TM do not follow diverse content, and require human domain-specific knowledge. The generation of vegetation is probably the most successful case of video game PCG through Traditional Methods, with the development of tools such as SpeedTree~\cite{SpeedTree}. The usage of SpeedTree in games, both AAA games and indie games, is widespread. SpeedTree was first launched in 2002, and since then there has been no similar successful tool. Traditional methods have mostly been used for offline generation; we found no references to TM online generation. The fact that they have been restricted to offline use hinders their use for replay-ability\footnote{The potential for continuing playing after the first completion of a game.}.

Machine Learning methods train models to generate content based on training data, thus reducing the need for human domain-specific knowledge. ML online generation is still not widespread, most of the work on ML is offline generation. The use of ML methods for PCG (PCGML) was recently reviewed in 2018 by Summerville \etal~\cite{Summerville2018}, who could not identify any successful case study used by the industry. One of the open challenges of using PCGML is the lack of training content, which is not always accessible, thus requiring additional effort from the development team to create the training content in advance. While the use of Deep Learning (a subfield of ML) has provided some advances in PCGML as reviewed by Liu \etal~\cite{liu2021deep}, its application to PCG is still limited by two main factors. First, DL approaches resist the specification and enforcement of explicit constraints, such as setting up the number of rooms in a dungeon. Constraints are important for game developers to achieve their vision of the content. Furthermore, DL has issues with the interpretation of results~\cite{puiutta2020explainable}, making it difficult for developers to understand the design patterns that appear in the generated content. The understandability of the patterns beneath the content is important for the development, since it allows developers to generate the content exactly as expected in the design. This is perhaps less of a problem in AAA and indie games, where creativity is encouraged to a certain degree, but a major issue for serious games.

Search-based methods do not suffer from the limitations of TM, and are therefore less expensive to apply for many types of content. SB methods also do not suffer from the limitations of ML methods, since they are easier to constrain and provide more accessible explanations of the generated results. In fact, some of the weaknesses of other methods become a strength in SB methods. For instance, in SB methods, the objective function can use constraints to guide the search. 
SB approaches for PCG generate content that is evaluated within the algorithm prior to its use. During the evaluation phase, content is appraised to decide whether to use it, discard it, or recombine its building blocks to generate further content. 

The scope of this survey puts the focus on SB methods, since they mitigate or outright remove a series of limitations suffered by TM and ML methods alike. 
In 2010, Togelius \etal~\cite{Togelius2011} surveyed SB methods for PCG (SBPCG) reviewing the very first research work in this field. They identified several research challenges for SB methods: Suitability of the generated content, avoidance of catastrophic failure, and improvement of the key components of the SB methods.

Our study reveals that in 2023 many of those challenges are still open, whereas new types of contents and concepts have appeared such as surprise search~\cite{gravina2016constrained} or quality diversity~\cite{withington2020illuminating}. Thus, making this field an exciting avenue for future research.

In this paper we survey the current state of SBPCG by analysing a total of 118 articles, published between 2010 and 2022, that  have applied at least one Search-Based method to procedural content generation. Based on the analysis of this article we evolve previous taxonomies~\cite{Togelius2011,Hendrikx2013} in a new one, which is able to capture the most recent content proposed in the literature, we describe the current work according to this taxonomy, and conclude by providing insights for future directions in this field.
\section{Background}
This section provides the reader with an overview of PCG and SBPCG.

\subsection{Procedural Content Generation}
PCG refers to the automation or semi-automation of the generation of content in video games. By content, we refer to every aspect of a game. This definition is broad given the large amount of content that a game usually needs, starting from the environment till the inner system logic  of the game. To that extent, the literature groups the content in different content types~\cite{Hendrikx2013}; game bits, game space, game system, game scenarios, game design, and derived content. 
Game bits, game space, game system, game scenarios, and game design, refer to elements inside the game such as vegetation or sound (game bits), environment or terrain (game space), Non-Playable Characters (game system), levels or puzzles (game scenarios), and rules or restrictions (game design). Derived content, on the other hand, is all the content that are generated because of a game, like videos of player experiences playing a game.

As discussed in the Introduction, the automated content creation can be achieved  through the application of several different methods that can be grouped in three main categories~\cite{Barriga2019} (namely Traditional , Machine Learning, and Search-Based methods). Our work surveys the current state of Search-Based Procedural Content Generation research by analysing new types of content and concepts, and providing insights for future directions in this field. In the following subsection we provide the reader with some background knowledge on SBPCG.

\subsection{Search-Based Procedural Content Generation}

\begin{table*}[t]
    
    \caption{Definitions of the key components for Search-Based Procedural Content Generation (SBPGC).}
    \resizebox{0.7\columnwidth}{!}{
    \begin{tabular}{ p{0.2\textwidth}  p{0.2\textwidth}  p{0.52\textwidth} }
    \hline
    & \textbf{Name}                   & \textbf{Short Description}                                                                                                            \\ \hline
    \multicolumn{1}{l}{\multirow{2}{*}{Encoding}} & Direct                 & Less genotype-to-phenotype complexity conversion                                                                             \\ \cline{2-3} 
    
    \multicolumn{1}{c}{}                          & Indirect               & Requires human effort on creating the conversion system genotype-to-phenotype                                                \\ \hline
    \multirow{6}{*}{Objective Function}             & Direct - Theory Driven & The developers assess with their opinion to elaborate the objective function                                                 \\
                                                    & Direct - Data Driven   & The objective function is based on information about relevant parameters extracted from artefacts                            \\ \cline{2-3} 
                                                    & Simulation - Static    & The simulator agent does not change during the simulation                                                                    \\
                                                    & Simulation - Dynamic   & The simulator agent learns during the simulation                                                                             \\ \cline{2-3} 
                                                    & Interactive - Implicit & Players are outright asked for their opinions                                                                                \\ 
                                                    & Interactive - Explicit & Data is indirectly extracted or inferred from the observation of the actions of the players and the results of those actions \\ \hline
    \end{tabular}
}
    \label{tab:pcg_definitions}
    \end{table*}
    
Search-Based Procedural Content Generation is a field of the more general Search-Based Software Engineering (SBSE) research area~\cite{harman2001search, harman2012search} . 

SBSE seeks to reformulate Software Engineering problems as `search problems'. Given a search space of a particular problem, a search-based approach can look for an optimal or near optimal solution within a set of candidate solutions. 
The next paragraphs describe the most commonly used algorithms in SBSE (which are also summarised in Table~\ref{tab:pcg_definitions}), and how the two key ingredients of SBSE (namely representation and objective function) have been being applied in SBPCG.

Within the main algorithms used for SBSE, we find local search algorithms, single-objective evolutionary algorithms, and multi-objective evolutionary algorithms~\cite{harman2001search, harman2012search}. Local search algorithms receive a set of candidate solutions as input, and then determine one of the candidate solutions as the starting point for the search. In order to perform the search, they then iteratively move to a neighbour solution which slightly differs from the current solution and evaluate the new candidate solution according to the fitness function. Single-objective evolutionary algorithms use mechanisms inspired by the Darwinian concept of evolution. Evolutionary algorithms apply genetic operations such as the crossover or mutation of individuals over the candidate solutions to obtain new populations of candidate solutions, which are then evaluated according to a fitness function that targets a single objective. Multi-objective evolutionary algorithms work through the same Darwinian principles and genetic operations to search for the best candidate within a search space, but evaluate the candidate populations according to objective functions that consider more than one goal.

A search-based approach needs a representation of the particular problem that an algorithm can understand to perform the search (\ie encoding).
Regarding the representation of the problem, we find two main components, the \textit{genotype} and the \textit{phenotype}. The \textit{genotype} is the data structure that the algorithm will process, and the \textit{phenotype} is the data structure that will handle the evaluation part of the search. In other words, a phenotype is a conversion from a genotype. Based on the difficulty of the conversion from genotype to phenotype, we refer to `direct encoding' or  `indirect encoding'. A direct encoding implies less genotype-to-phenotype complexity conversion than an indirect encoding, which requires a more complex conversion system. While an indirect encoding requires more human effort on creating the conversion system, it also provides freedom to represent the content. In SBPCG, an example of a direct encoding is a grid where each cell represents an element of the content. An evolutionary algorithm is capable of interpreting the grid and evolve it, and the final phenotype can be extrapolated just by looking at the genotype. An example of indirect encoding is a list of the details of the different elements that compose the content. A vector representing that list of details is likely to be evolved by an evolutionary algorithm. However, there is a need for interpreting the details of the elements in the vector to generate the phenotype.

Finally, a search-based approach needs an objective function (or fitness function) that guides the search towards an optimal solution.

Regarding the objective function, SBPCG differentiates between three different types~\cite{Togelius2011}: direct, simulation, and interactive. Direct objective functions are those that are based on the available knowledge of developers (that is, the developers themselves participate in the assessment of the objective function). Direct objective functions can be either theory-driven (meaning that the opinion of the developers is directly leveraged) or data-driven (meaning that information about relevant parameters is extracted from artefacts like questionnaires or player models). Simulation objective functions replicate real situations to estimate the behaviour of real players. Work in this area focuses mainly on developing more human-like agents, bots, and AIs to be used by objective functions. Simulation objective functions can be static, where the simulator agent does not change during the simulation, or dynamic, where agents that learn during simulation are used. Finally, interactive objective functions are those that involve players in the composition of the objective function. Incorporating human expertise in the objective function constitutes a broad research area itself, named human-in-the-loop~\cite{wu2022survey,kruse2022evaluation}. 
In SBPCG, interactive objective functions can be either explicit, when players are outright asked for their opinions, or implicit, when the data is indirectly extracted or inferred from the observation of the actions of the players and the results of those actions.
\section{Survey Methodology}
\label{sec:methodology}

This survey gathers and categorizes research work published in the field of PCG for Game Software Engineering. More precisely, this work puts the focus on the context of SBPCG. In the following subsections, we present our search methodology in detail, along with a description on the selected publications.

\subsection{Search Methodology}
Inspired by Martin \etal~\cite{martin2016survey}, our search methodology follows three steps: First we perform a preliminary search followed by a repository search, then we apply the selection criteria, and finally, we conduct a snowballing process.

Our preliminary search has two goals. The first one is to assess whether there is a sufficient amount of publications in this field since the latest survey on Search-Based PCG~\cite{Togelius2011} from 2011. The second one is to define the keywords that will be used for the repository search.
The results of the preliminary search define the following keywords: ( `pcg' OR `automatic generation' OR `procedural' ) AND ( `videogame' OR `game' ) AND ( `search-based' OR `evolutionary' OR `genetic' OR `local search' OR `tabu search' OR `Monte Carlo Tree Search' OR `mcts' )

We conduct the repository search on Scopus, a database that indexes papers from ACM, IEEE, Springer and Elsevier, among others. We have gathered publications from 2011 to 2022, since the last survey on Search-Based PCG~\cite{Togelius2011} released on 2011, which covers work published until 2010. We restrict the publications based on the words defined during the preliminary search. We run the query on the title of the article, the abstract and the keywords associated to each publication.

The inclusion criteria used in this survey ensure that the publications present Search-Based algorithms with the aim of generating content in the area of video games. For example hybrid approaches such as Neuroevolution, that is, the use of an evolutionary algorithm to evolve a neural network (or any other ML method) that will create the content, are not included. Similarly, approaches that evolve other agents assessing content rather than creating it, are not included either.
To verify that the publications found during our search meet the inclusion criteria, we examine the publications by applying the following three stages process:

\begin{enumerate}
  \item Title: we remove publications that are clearly irrelevant from the title.
  \item Abstract: we inspect the abstract and remove publications that are clearly irrelevant according to the scope defined in Section~\ref{subsec:scope}.
  \item Body: publications that pass the previous two steps are excluded if their content is not relevant to the scope of this survey )(Section~\ref{subsec:scope})
\end{enumerate}

The publications studied in this survey are the result of the application of this selection process. Sections~\ref{sec:game_bits} to~\ref{sec:game_design} discuss in detail the studied publications.

To reduce the risk of missing relevant publications from the literature, we apply one level of backwards snowballing~\cite{WohlinClaes2014Gfsi}. More precisely, we inspect the publications cited in the related work of each publication that passed the previous selection process. 

In addition, we request feedback from the authors of the work included in our survey, as done in previous surveys~\cite{martin2016survey}.

\subsection{Selected Publications}

\begin{table}
  \caption{Number of publications retrieved at each step of our literature search.}
  \resizebox{0.65\columnwidth}{!}{
  \begin{tabular}{p{0.4\linewidth}p{0.23\linewidth}p{0.23\linewidth}}
    \hline
    \textbf{Step}&\textbf{Number of publications}& \textbf{Added publications}\\ \hline
    Preliminary search&301&104\\ \hline
    Snowballing&196&13\\ \hline
    Author feedback&1&1 \\ \hline
  \end{tabular}  
}
  \label{tab:methodology}
\end{table}

Table~\ref{tab:methodology} provides the number of publications we retrieved at each stage of our search, and specifies the number articles that at each step meet the inclusion criteria for this survey, and were therefore included in the survey.  

As a result of the examination of 498  publications, we retained 118 unique publications that are in the scope of our survey, i.e., articles describing  content generation for video games by applying at least one SB technique. 
These 118 publications have been published in 43 different venues. The list is available in our online appendix~\footnote{\url{https://solar.cs.ucl.ac.uk/os/sbpcg/Venues.xlsx}}, along with a classification of the publication venues according to the CORE ranking portal~\footnote{\url{https://www.core.edu.au/conference-portal}}, and the JCR ranking portal~\footnote{\url{https://jcr.clarivate.com/jcr/home}}.

\subsection{Threats to Validity}
To mitigate the threat of missing relevant information in our literature survey we undertook a number of mitigation actions, as detailed below.    0po-Two authors examined all articles independently, in order to ensure reliability and reduce researcher bias. The results were compared at the end of the process, and any inconsistency was resolved by a joint analysis and discussion. 
Moreover, to ensure that our survey is comprehensive and accurate, we contacted the authors of the publications collected. We asked them to check whether our description about their work is correct. Based on their feedback, we revise our survey as well as included 1 additional publication. 
We carefully describe the search process we followed and make additional data available in our online appendix, so that future studies can reproduce, replicate, and extend our work. 
\section{SBPCG Taxonomy}
\label{sec:taxonomy}

 The main aim of creating an SBPCG taxonomy is to correctly identifying and classify the different type of contents that have been described in published work on SBPCG. This taxonomy is then used to discuss each of the articles within a given category. This will ease the analysis and comparison of work aiming at automatically generating new type of content. In this section, we explain the methodology by which we construct the taxonomy that we use to classify existing SBPCG work.

\begin{figure*}[t]  
        \centering  
        \caption{Summary of the process we followed to construct a taxonomy of procedural content categories. We started from two existing taxonomies (\cite{Hendrikx2013, Togelius2011}, and after a first Harmonization step, we proceed to extend them in three additional steps based on the analysis of the articles included in this survey. A detailed description of each step is provided in Section \ref{sec:taxonomy}.}
        \includegraphics[width=0.7\textwidth]{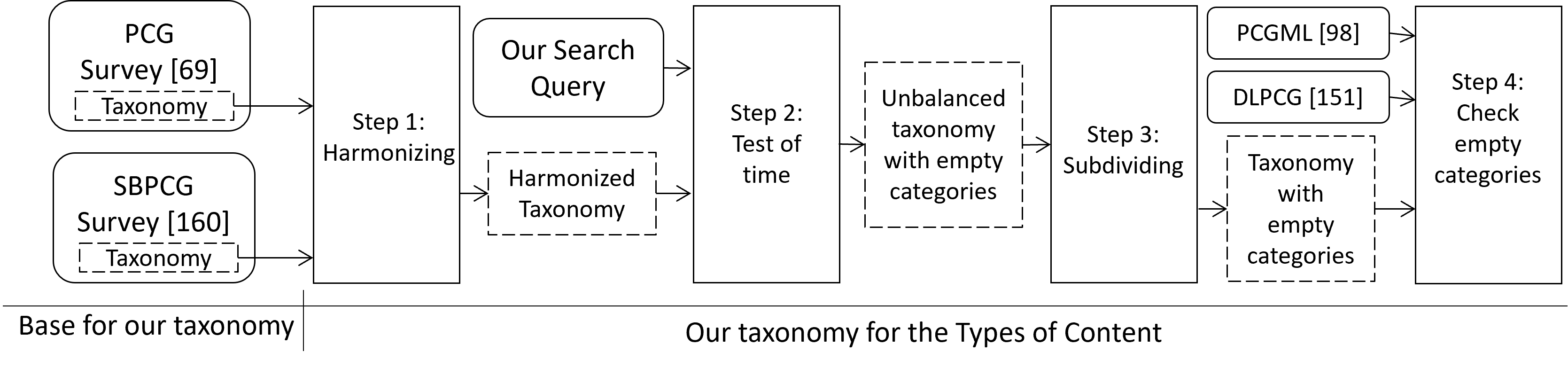}
        \label{fig:taxonomy}
    \end{figure*}

Figure~\ref{fig:taxonomy} summarizes the steps we followed elaborating the taxonomy. In Step 1, \textit{Harmonizing}, we analysed the taxonomies from two previous studies and looked for similarities, thus getting a starting point for an harmonized taxonomy. Then in Step 2, \textit{Test of time}, we create a preliminary categorisation of the work that we analyse in our survey. We then expand this categorization by including additional subcategories  to better reflect the current state of the work in SBPCG in Step 3, \textit{Subdividing}. Finally, in Step 4, \textit{Check Empty Categories}, we analysed why some types of contents that was discussed in the previous surveys, have not been subsequently get any traction in SBPCG, thereby leading us to the removal of some categories. Further details for each of the steps are provided in the following paragraphs.

Previous PCG surveys have adopted a similar approach, by proposing their own taxonomy.  Our taxonomy stems from the analysis of these  previous work, and also includes new types of content according to the needs of newer articles published from 2011 to 2022.
Specifically, we used two surveys (namely Togelius \etal~\cite{Togelius2011}  and Hendrikx \etal~\cite{Hendrikx2013}) as the starting point for our taxonomy as they are more generic and cover more types of content than other surveys~\cite{Summerville2018,liu2021deep}. Togelius \etal~\cite{Togelius2011} focused on Search-Based techniques (as we do) proposed till 2011; while the 2013 survey by Hendrikx \etal~\cite{Hendrikx2013} focused on the most common and some emerging techniques for PCG (including 11 Search-Based ones). Table~\ref{tab:taxonomies} summarise and compare the two previous surveys and our own by specifying their publication venue, the publication time frame of the articles discussed by each survey, and their main focus\footnote{These information have been extrapolated from the content of the surveys, as they are not explicitly discussed in the surveys.}.
 
 In the following paragraphs, we explain the taxonomies used in both the previous surveys, and the process we followed to derive the one we use in this survey.
 
Togelius \etal~\cite{Togelius2011} divided SBPCG content into two categories: considering whether the content was \textit{necessary}, that is, whether the content was needed for a game to be played, or \textit{optional}, that is, whether the content was not strictly needed for the game to be played and could be avoided during gameplay. The types of content on each category may vary depending on their purpose in a game. This survey built the taxonomy considering research articles published until 2010.

Hendrikx \etal~\cite{Hendrikx2013} proposed a more structured classification for PCG methods. Instead of constructing their taxonomy from the results of a literature search, they asked themselves about the contents of a game. Then, they looked into the different types of content to analyse the techniques applied within each category. 

\begin{table*}[]
    \centering
    \caption{Summary of the three surveys.}
    \resizebox{0.6\columnwidth}{!}{
    \begin{tabular}{lllrl} \hline
    Authors         & Published in               & Cover articles from     & Number of articles & Focus \\ \hline
    Togelius \etal~\cite{Togelius2011}    & ToG          & 2005 - 2010   & 25        & SBPCG \\ \hline

    Hendrikx \etal~\cite{Hendrikx2013}    & TOMM          & 1991 - 2011  & 80        & PCG \\ \hline

    This survey    &  \textit{under review}  & 2011 - 2022  & 118        & SBPCG \\ \hline
    \end{tabular}
    \label{tab:taxonomies}
}
    \end{table*}

We have studied the detailed descriptions of both taxonomies, appreciating significant similarities between them (see Fig.~\ref{fig:taxonomy} Step 1). Both taxonomies overlap in 10 out of 12 types of content of the sub-classification of the taxonomy of Togelius \etal~\cite{Togelius2011}. More precisely, there are three types of content with the same name (buildings, puzzles and levels) in both taxonomies, and we argue that vegetation subsumes trees, indoor maps subsumes maps, outdoor maps subsumes terrains, storytelling is equivalent to storyboards, story is equivalent to narrative, and system design subsume rules and mechanics. 

To build a first version of our taxonomy we start from the those of Hendrikx \etal. Although their focus is different from ours, their taxonomy is more comprehensive than those by Togelius \etal 's one, and we observe that it better fits the more recent articles we found in our search till 2022. We augment this taxonomy with three types of content from the taxonomy of Togelius \etal (\ie, weapons, tracks, and camera control) because we found recent articles that tackle those types of content. Weapons fit in game bits, tracks fit in game scenarios, and camera control fits in game design.

\begin{figure}   
    \centering
    \caption{Structure of our proposed taxonomy of procedural content categories. The subcategories highlighted in grey are those that have been introduced in this taxonomy based on the articles we found,  thus extending previous existing taxonomies.}
    \includegraphics[width=0.7\linewidth]{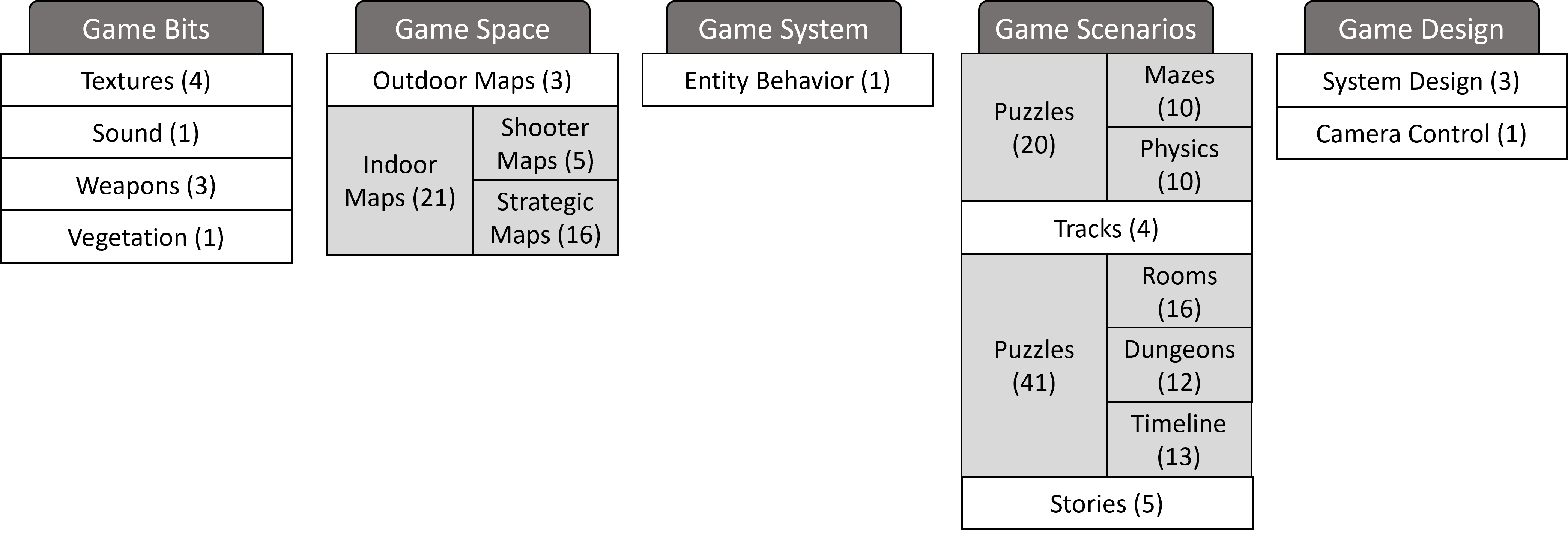}
    \label{fig:unbalance}
\end{figure}

Then, we analysed our literature search results and categorised the articles found according to this preliminary taxonomy (see Fig.~\ref{fig:taxonomy} Step 2). Such an analysis pointed out two aspects that must be taken into account in devising our new taxonomy: difference in the amount of work related to different type of content, and decaying vs. emerging type of content. 
Firstly, the number of work within the different types of content vary significantly (see Fig.~\ref{fig:taxonomy} Step 3). In particular, there are two types of content (levels and indoor maps) that independently comprise more work than those presented by the rest of the types of content combined. Looking into those types of content we have identified differences over the content that allow a further sub-categorization. We therefore propose a further sub-categorization of levels, indoor maps, and puzzles. To that extent, we have divided levels into the timeline, room, and dungeon subcategories; indoor maps into the shooter maps, strategic maps, and floor subcategories; and puzzles into the physics and mazes subcategories (see Fig.~\ref{fig:unbalance}). We argue that this new classification is better adapted to the current research in SBPCG.
Secondly, one of the general categories mentioned above (\ie derived content) and half of the types of content defined in the taxonomy subcategories have not been tackled in the last decade by SBPCG works. The derived content category addresses the generation of content after the development of a game. This type of content is usually related with the interaction of players with a game. Derived content is out of our scope as we relate to the generation of content for a game before its release or during its gameplay.

From the analysis of the literature we also noticed that there are types of content that have not been tackled by SBPCG. To speculate about the reason (see Fig.~\ref{fig:taxonomy} Step 4), we have examined two more recent surveys of PCG~\cite{Summerville2018,liu2021deep}, which tackle PCG from the point of view of PCGML. PCGML handles patterns better as it is based on learning, while SBPCG needs a manual configuration of the constraints in order to be able to follow patterns. Maybe there are types of content that require so much effort in the form of constraints that researchers resort on learning the constraints.

In the last ten years there are types of content that have not been tackled by neither SBPCG nor PCGML (\ie buildings, behaviour, elements, water, ecosystems, road networks, urban environments and world design). We ran an informal search in Google Scholar for each type of content, in order to understand these types of content, with the following query: "procedural generation" + $<$ type of content $>$. The results show that works that study these types of content are addressed through Traditional Methods, such as, noise generators, fractal structures or L-systems~\cite{SmelikRubenM2014ASoP}. Here we find an opportunity for SBPCG and PCGML. TM approaches have been successful but we did not find arguments proving their efficiency over SBPCG or PCGML techniques.

Further explanation about the general categories, types of content, and subcategories can be found within the following sections, which compose the survey \textit{per se}.  The Game Bits Section (Section~\ref{sec:game_bits}) addresses textures, sounds, weapons, and vegetation. The Game Space Section (Section~\ref{sec:game_space}) addresses indoor maps (and their subcategories into shooter maps, strategic maps, and floors) and outdoor maps. The Game Systems Section (Section~\ref{sec:game_systems}) addresses non-playable characters. The Game Scenarios Section (Section~\ref{sec:game_scenarios}) addresses puzzles (and their subcategories into mazes and physics), levels (and their subcategories into timeline, room, and dungeon), tracks, and stories. The Game Design Section (Section~\ref{sec:game_design}) addresses system design and camera control. In the online appendix \footnote{\url{https://solar.cs.ucl.ac.uk/os/sbpcg.html}}, we list the articles discussed in this survey by reporting on their publication year and venue, the two main characteristic of the search-based approach  (\ie encoding and objective function) and content type they investigate (see Figure~\ref{fig:unbalance}).
\section{Game Bits}
\label{sec:game_bits}

We start our survey by discussing Game Bits. Game bits are the smallest units of game content, or in other words, the most basic pieces that can be used to build a game. Game bits on their own, that is, when considered independently and out of the context of a particular game, do not hold any value for the players of the game. In that sense, game bits are the most basic building blocks that are in turn used by the developers of a game to generate other types of content. For instance, textures are one of many different game bits that can be used to construct game scenarios. We analyse work with respect to textures, sounds, weapons and vegetation. Textures are the images and materials of the elements of a game. Textures are in accordance with the artistic style of the game.
Sounds encompass the music and sound effects of a game. Music is an important element to create the game atmosphere.  Sound effects report feedback to the player regarding their actions or changes in the environment. 
Weapons  are game bits that are used by the players to face adversities in a game.
Vegetation creates an aesthetic engaging environment. In addition, this game bit help players as hiding place, as raw material, or guide players about directions and climate changes.

\subsection{Texture}

Textures have been mainly addressed by the Graphical Computation community~\cite{sims1991artificial}. The generation of textures for video games tackle a wide range of challenges. For instance, there are work that generate the shape of the elements of a game~\cite{kowalski2018mapping}. Kowalski \etal~\cite{kowalski2018mapping} generated novel shapes for chess-like games motivated by previous work related to chess-like games that used the rules of chess to generate novel games (see Section~\ref{subsec:system_design}). Their work tackled the generation of the whole collection of pieces for a game, as well as each individual piece separately. The evaluation indicated that keeping the balance between the collection and the individuals moved the results towards one of the objectives.

Other work have tackled the generation of textures by evolving the colour palette~\cite{liapis2018recomposing} or the complex materials~\cite{bernardi2020procedural} that compose an element of a game. Players tend to associate the appearance of an element of a game with its aim, that means, that a change in the appearance of an element of a game would affect the perspective of the player. Liapis~\cite{liapis2018recomposing} studied the use of an evolutionary approach to modify the colour palette of Pokémon keeping their shape. The new colour would fit those Pokémon into different Pokémon categories just by their appearance.

Brown \etal~\cite{brown2022evolving} evolves camouflages patterns for game assets. Inspired by a real military uniform pattern, a genetic algorithm is able to generate new patterns assessed by the environment. The evaluation used computational vision observation of the pattern in the environment, and assessed its capacity to camouflages. They also conducted a human evaluation that corroborated the blend capacity assessed previously by a computer.

\begin{mdframed} 
    The work on textures shows how the search in a large space leads to provide new ideas over this type of content, and the human evaluation corroborated how new content would be feasible in a video game.
\end{mdframed}  

\subsection{Sound}

The application of algorithms for sound composition tasks is not a new challenge. Research in algorithmic composition started in the last decade and has a long history~\cite{jarvelainen2000algorithmic}. Its application on games is not new either~\cite{collins2009introduction}. After the survey of Togelius \etal~\cite{Togelius2011}, Plans \etal~\cite{plans2012experience} brought Experience-Driven PCG and sound together. Plans \etal generated music based on the experience of players while playing. The actions of players are the inputs of the approach, and affect the music at runtime. Their results sustained the idea that music affects gameplay.

\begin{mdframed} 
    A key factor over the sounds is the effect on the players, generating new sound content seems to require some human verification. 
    This may be the reason why there is not so much SBPCG work on this type of content.
\end{mdframed} 

\subsection{Weapons}

Previous work on procedural generation of weapons tackled the challenge through approaches guided by human players~\cite{hastings2009evolving,hastings2009automatic,hastings2010interactive}. The authors of these work developed a 2D commercial game where the weapons were generated based on the interaction of the players with the game. More precisely, the game generated weapons based on how often the players interacted with a weapon in particular.

McDuffe \etal~\cite{mcduffee2013team} also applied an interactive objective function for generating weapons. In contrast to the work referenced in the prior paragraph, McDuffe \etal generated weapons for an academic 3D multiplayer game. The weapons were evolved through the study of implicit evaluations of each weapon provided by players. The measured factors were the time that players had equipped each weapon and the number of kills obtained by players with each weapon. From the point of view of the developers, the results suggested that the approach generated interesting weapons.

Inspired by the work of McDuffe \etal~\cite{mcduffee2013team}, Gravina \etal~\cite{gravina2015procedural} introduced a weapon generation approach for a 3D commercial game. They evolved weapons with the aim of obtaining balanced weapons, where balance was calculated through an objective function that contemplated the distribution of kills of each weapon. In addition, they applied the same objective function to evolve and improve existing weapons. In  contrast to the work by McDuffe \etal~\cite{mcduffee2013team}, the objective function simulated matches among bots, which did not need rendering to do the simulation thus accelerating the evaluation. The authors also ran an experiment with human players to assess the quality of the generated weapons. The received feedback showed that the approach generated weapons that were interesting, fun to play, and balanced.

Based on their previous work, Gravina \etal~\cite{gravina2016constrained} continued with the generation of balanced and effective weapons, where balance still considered the distribution of kills of a weapon and effectiveness considered whether a weapon could actually kill an enemy or not. Throughout the paper, they explored the usage of a constrained surprise search approach to generate weapons. Constrained surprise search is a Feasible Infeasible-2population constrained optimization algorithm which looks for ``surprising'' results, that is, diversity on the output. Gravina \etal~\cite{gravina2016constrained}  compared the performance of their approach against a single-objective search approach and a constrained random search approach. The comparison exposed that the objective search approach obtained more feasible results than the other two approaches. However, the surprise search approach obtained more diversity in the results than the other two approaches. Overall, the results of this work highlighted that the constrained surprise search approach was capable of quickly and reliably generating diverse weapons.

\begin{mdframed} 
    We observe a successful use of on-line generation for weapons where the search is guided through players feedback to create the content that would suit those players. 
    On the other hand, weapons generation has moved also towards generating diverse content that could be unexpected and not only suiting players preferences.
\end{mdframed} 

\subsection{Vegetation}
Vegetation includes wherever plants appear in any digital environment, including indoor plants. Vegetable, like any digital object, can be interactive or not.  To date, almost all vegetation is noninteractive. The importance of detail representation of vegetation, depends heavily on the target realism of the game and the resolution of the hardware. Empirically, vegetation is well-suited for TM:  certainly, the game industry has overwhelming voted for it with their keyboards for it to date. Evidently, the industry deems the coarse-grained realisation of vegetation satisfactory. 
The canonical TM is SpeedTree, which provides a development environment to create and modify vegetation, as well as a wide catalogue. The development environment allows the user to visualize the vegetation in different seasons, and to add wind, and light, among the different options. SpeedTree is widely used in cinema and video games, two of the most popular 3D game engines~\cite{toftedahl2019taxonomy} (Unity and Unreal) integrate it.
Perhaps, TM's success has stymied ML work in the space;  in any case, we are unaware of any PCGML for vegetation.

The uniformity of TM vegetation, TM's weakness, inspired one group of SBPCG researchers to look to improve the diversity of noninteractive vegetation.
Mora \etal~\cite{mora2021flora} has proposed a novel, SBPCG approach to generate vegetation. 
Their approach uses an evolutionary algorithm to simulate the life cycle of the flora. 
Universe 51 game is a planetary exploration game, one with photo realistic environments which enhance the fun and interest of exploring an alien landscape.
The authors integrated their flora generator into Universe 51 and play-tested it.
Their goal is to increase the naturalness of a digital environment by simulating change in the flora.
There are two big questions unaddressed by Mora \etal, the computational cost of adding this dynamism, and the benefit in terms of player satisfaction.
Perhaps this is why, at the time of writing, no game publisher has adopted it.

Interactive flora is even less explored. To our knowledge, there is only one game that does so. Petalz~\cite{risi2015petalz}, a game which allows players to interact with flowers creating new content. Interactive objects, by definition, as noted in the introduction, have behaviours, which, in turn, introduce constraints, like weight, that are well-suited for SBPCG.

\begin{mdframed} 
    Flora generation has been dominated by TM techniques, however there is no evidence that prove why other techniques have not gained a similar widespread. 
    Answering the two open questions that we have aroused could enlighten the reason.
\end{mdframed} 
\section{Game Space}
\label{sec:game_space}
Game spaces (or maps) are the environments of a game, that is, a one-, two-, or three-dimensional area that can be filled with game bits in a relative position and direction. Game spaces do not specify linear gameplay, meaning that they do not need a start and end point. We can distinguish two main  types of game spaces: outdoor and indoor maps.  Outdoor maps are large spaces, usually with different ecosystems, that require a certain amount of time to traverse. In these spaces it is common to use vehicles or teleport systems. Specific to outdoor maps is the topography of their terrain, which is a depiction of the elevation of the map. Most works in the field of outdoor maps deal with the topography of the terrain of the map, and often referred to it just as terrain.
Indoor maps are maps that are contained in a limited space within which the player can move. For example, shooter maps rely on buildings and objects to create suitable combat spaces, while strategic maps are designed to require the management of different kinds of assets in real-time. The assets are the different kinds of units, resources, buildings, and any other video game elements that a player can position, manoeuvrer, manage, or otherwise control during play.      

\subsection{Outdoor Maps (Terrains)}

In 2009, Frade \etal~\cite{frade2009breeding} coined the term Genetic Terrain Programming (GTP), which refers to the use of SBPCG in order to generate terrain for video games. Their approach used an interactive objective function that involved humans to guide the search. The results highlighted two limitations regarding user fatigue and the inability to perform zoom over the generated pieces of terrain. 

In order to address these limitations, Frade \etal continued their work with a new version of GTP, named Automated Genetic Terrain Programming (GTPa)~\cite{frade2010aevolution,frade2010bevolution}. These works modified their previous approach to tackle these limitations by avoiding user fatigue and enabling the zoom feature over the generated terrain. In order to avoid user fatigue, the authors proposed two distinct direct objective functions, guided by different objectives in each of the two works, that did not involve humans. In 2012, the same authors combined the two functions from their GTPa previous work into a new direct objective function that was the result of the sum of the formerly individual objectives~\cite{frade2012automatic}. The approach decreased overlaps in the results, but generated terrain with smaller amplitude. The results were released in an open database\footnote{\url{https://sourceforge.net/projects/tps-db/}} for future research purposes~\cite{frade2012aesthetic}. 

In 2016, Pech \etal~\cite{pech2016using} proposed a novel approach for generating terrains by incorporating elements into pre-existent terrains. To incorporate those elements, Pech \etal~\cite{pech2016using} introduced the use of an architectural element in the objective function, which is isovist. An isovist is the volume of a space visible from a given point in space. Through this perspective, Pech \etal~\cite{pech2016using} were able to introduce elements such as hidden areas in a terrain. This novel approach benefited terrains that were previously unplayable.

\begin{mdframed} 
   Terrains generation has evolved from the use of an interactive objective function to a direct objective function to avoid user fatigue. It is interesting to notice that there is no comparison between the use of an interactive objective function or a direct objective function, and it gives an opportunity to explore simulation-based objective function to address the issue of user fatigue.
\end{mdframed}

\subsection{Indoor Maps}

\subsubsection{Shooter Maps}

The previous SBPCG survey~\cite{Togelius2011} shows that SBPCG work addressed the challenges of generating content as tracks for racing games, rules for board games, weapons for space shooters, levels for platform games and maps for real-time strategy games. The work of Cardamone \etal~\cite{cardamone2011evolving} is the first to address the challenge of generating First Person Shooter (FPS) maps. Despite FPS is one of the most popular game genres, only three studies~\cite{cole2004using,gorman2006believability,van2009hierarchical} investigated FPS games before, by targeting the generation of a new form of content; that is, generating the behaviour of Non-Playable Characters.

Maps are the heart and soul of FPS games. Cardamone \etal~\cite{cardamone2011evolving} argue that generating FPS maps poses a bigger challenge than generating maps for other games. They also acknowledge that FPS maps should favor gameplays that reward skillful use of complex tactics, and force players to vary their tactics so they cannot use the same patent trick all the time to win.

Outside of PCG, researchers have analysed FPS maps and proposed design patterns for FPS maps~\cite{hullett2010design}. These design patterns might be useful for guiding the automated search of new FPS maps. Nevertheless, Cardamone \etal~\cite{cardamone2011evolving} do not leverage the former design patterns but rely on bots for the objective function instead.

The work of Cardamone \etal ~\cite{cardamone2011evolving} generated complete and playable FPS maps for Cube 2 \footnote{\url{http://sauerbraten.org}}, an open-source game from 2004. They learned that direct encoding works better than indirect encoding for this kind of map. However, they point out two main limitations of their work: (1) the dependency on the control logic of the default bots of the game, and (2) the lack of validation by players, that is, they evaluate the maps in terms of synthetic measures (fighting time and map space), but they do not claim that their measures correspond with human players’ judgments.

Three years later, Lanzi \etal~\cite{lanzi2014evolving} revisited the challenge of generating maps for FPS. Their work shares the encoding, the case study, and the synthetic evaluation with the work of Cardamone \etal~\cite{cardamone2011evolving}. Nevertheless, Lanzi \etal~\cite{lanzi2014evolving}  aimed for a different goal.  Whereas Cardamone \etal addressed fast-paced action maps, Lanzi \etal addressed match balanced maps. Balancing a match between two players is one of the seminal problems of video game development. To do so, Lanzi \etal~\cite{lanzi2014evolving}  proposed a novel objective function computed on the basis of the statistics collected from a simulated match between two bots. Their work succeeded, in three scenarios, generating maps that balanced the match between pairs of bots with different skills.

Olsted \etal~\cite{olsted2015interactive} brought human players to the work of Cardamone \etal~\cite{cardamone2011evolving}. The novelty of their work was the use of human players as objective function. Humans played the maps and used votes to rank them. The approach was evaluated in an academic context with the FPSEvolver video game, a video game developed by the authors for the evaluation. Almost every player agreed that the maps improved in quality as they played. However, players expressed that maps felt quite flat in comparison to maps of real games such as the popular Counter-Strike and Call of Duty.

Loiacono \etal~\cite{loiacono2017fight,loiacono2018multiobjective} were the first to explore multi-objective algorithms for generating FPS maps. As in previous work, they use the same encoding as Cardamone \etal~\cite{cardamone2011evolving} and Lanzi \etal~\cite{lanzi2014evolving} (\ie a static simulation through bots to guide the search) and the same case study (\ie Cube 2). However, Loiacono \etal~\cite{loiacono2018multiobjective} collect statistics from the simulation for multi-objective search. Their objectives are the balance of the map, the pacing on which the players are engaged in fights, the average length of kill streaks, the fighting time, the shooting time, ability of loose enemy's sight during fight and loose enemy's sight enough time to stop the fight. Their evaluation suggest that multi-objective evolution can provide a good insight of what happen with human players. The same objectives computed using bots and evaluated with human players provide significant agreement.

\begin{mdframed} 
  Shooter map generation approaches have addressed several challenges, starting with the use of single and multi objective search algorithms. They have also used simulation and interactive objective functions. Finally, their evaluations used both academic and commercial video games. One of the works unveil players concern that the content felt quite flat, and this issue seems to remain unaddressed in the literature.
\end{mdframed}

\subsubsection{Strategic Maps}

In the previous SBPCG survey~\cite{Togelius2011}, it is possible to appreciate that SBPCG works tackled map generation for strategic games. Those work generated strategic maps at the scale of academic games~\cite{togelius2010towards,togelius2010multiobjective}. Togelius \etal~\cite{togelius2010towards,togelius2010multiobjective}, authors of those works, used a semi-direct encoding, crossover and mutation operators, and a five-objective function. The objectives were obtained individually for each game, and were derived from the play style and rules of each specific game under evaluation. Four of these objectives (surface, asymmetry, resource distance, and resource clustering) were of the direct type, and the fifth (A* base distance) was of the simulation type. In 2013, Togelius \etal~\cite{togelius2013controllable} extended their previous work with further experiments and included human players in the evaluation. Human players appreciated that the asymmetry objective generated unbalanced maps.

In the last ten years, SBPCG moved from academic strategic games to commercial strategic games. Lara-Cabrera \etal~\cite{lara2012procedural} were the first that generated strategic maps at the scale of commercial games (in particular, they did so for Planet Wars, an indie game). To do that, they went beyond the state of the art by leveraging bot-based simulations to guide the search~\cite{lara2012procedural,lara2013procedural}. Specifically, they identified the indicators that should be monitored during the simulation to calculate the objective function~\cite{lara2013using,lara2014balance}. Those indicators, once again specific to the game being evaluated, addressed map balancing (territorial imbalance, growth imbalance, and enemy imbalance), resource management dynamism (game length, conquering rate, reconquering rate, and peak difference), and player confrontations (battle rate and destroyed enemies). These works claimed that they successfully generated strategic maps, however, none of them involved humans in the evaluation to evaluate the quality of the results or to find out whether the results were aligned with the expectations of the players.

In 2013, Lara-Cabrera \etal continued their work on strategic maps from a different perspective: The aesthetics of strategic maps~\cite{lara2013evolving,lara2014self,lara2014geometrical}, that is, the spatial distribution of the elements of the map and their features (size and number of elements). In contrast to their previous work, the authors based their research on direct objectives through which they assessed the geometric, morphological, and topological properties of the maps with the purpose of generating procedural content related to aesthetic aspects of the game. The results of these works were evaluated through the usage of automated and semi-automated techniques along with the support of a human expert. Contrary to previous work~\cite{lara2013using,lara2014balance}, the evaluation of these approaches intended to measure the degree of the quality of the aesthetics of the maps according to the defined objectives, with the aim of studying whether the generated maps were aligned with the expectations of the players.

The aesthetics of strategic maps have also been addressed by other authors before. Through a constructive method, Johnson \etal~\cite{johnson2010cellular} used a Cellular Automata (CA) algorithm for generating maps. A CA algorithm is a discrete model with self-organizing properties that consists of a grid containing cells that can exist within a finite number of states. The algorithm works by setting a state in each cell of the grid and traversing the grid through an iterative process.
However, CA algorithms lack control and cannot be easily adapted for generating other maps. For this reason, Mahlmann \etal~\cite{mahlmann2012spicing} generated maps using a search-based approach that incorporated a control mechanism to the CA algorithm approach. To that extent, they use a direct objective function that generates maps for an abstract version of another strategic game (Dune 2).

The most recent work by Lara-Cabrera \etal~\cite{lara2017procedural} is the first one that tackled level balancing for a 3D academic strategic game (\ie Paintbol). The authors defined a balanced level as a level that does not provide an initial advantage for one of the two participating teams over the opposing team. In contrast to their previous work on balancing content~\cite{lara2012procedural} (with a simulation objective function), Lara-Cabrera \etal~\cite{lara2017procedural} used an indirect encoding and a direct objective function that analysed the defensiveness, ranking, and dispersion indicators. Those indicators generated balanced levels based on the results of the objective function. The results of this work suggested that, by applying different parameters in the evolutionary algorithm, better maps were generated. In particular, the rank and elitist selection methods generated better maps than roulette selection according to the measured indicators.

Other authors also tackled the generation of balanced maps~\cite{barros2015balanced,kowalski2018strategic,franco2022generating,ma2021angle}. Those approaches used search-based techniques to obtain the balanced maps rather than generating them from scratch. Barros \etal~\cite{barros2015balanced} balanced the maps through the initial position of the players. Kowalski \etal~\cite{kowalski2018strategic} and Franco \etal~\cite{franco2022generating}  evolved the positions of the assets that would be placed in the map. Ma \etal~\cite{ma2021angle} also placed the assets in the map, but using a multi-objective approach The results by Barros \etal~\cite{barros2015balanced}, by Kowalski \etal~\cite{kowalski2018strategic} , by Franco \etal~\cite{franco2022generating}, and by Ma \etal~\cite{ma2021angle} generated playable and balanced maps.

\begin{mdframed} 
  One of the main goals in strategic map generation is the balance in the maps generated. Several approaches have addressed this goal from different perspectives (e.g., different representations and objective functions) but only one work has provided a human evaluation that concludes that asymmetry does not work for balanced maps. The other main goal addressed by strategic map generation is based on the aesthetics of the maps. We find it interesting that contrary to the balance goal, researchers have taken into account human expectations but they have only investigated the use of direct objective functions.
\end{mdframed}
\section{Game Systems}
\label{sec:game_systems}

Game systems bring the virtual worlds of video games closer to the human world in order to provide the players with a sense of immersion. This is achieved through complex models that include, among others, entity behaviours, also known as Non-Playable Characters or NPCs. NPCs are essential for the experience of the players, since they generate the illusion of a virtual world along with the opportunity to create interactions between the environment and the players. 

In 2013, Guarneri \etal~\cite{guarneri2013golem} described an approach to automatic generate NPC monsters through an evolutionary algorithm. The goal of the approach was to obtain a diversity set of new monsters from a starting population defined by the developer. This approach was later applied by Norton \etal~\cite{norton2017monsters} on another video game genre. With the same goal, Ripamonti \etal~\cite{ripamonti2021dragon} developed a novel approach to generate monsters adapted to players. This approach records the number of times a player kill each type of monster, considering the monster with more death rate the preferred by the player. The evaluation used a simulation to test the generated monsters, meeting authors' expectations on diversity, coherence, and difficulty.

Pereira \etal~\cite{pereira2021procedural_enemies} instead of diversity seek for generating enemies that meet a difficulty criteria. In that sense, the objective function looks for enemies that are closer to the difficulty stated in the search. The results with human players indicates that the generated content matched the desired difficulty.
Viana \etal~\cite{viana2022illuminating} extended the work by Pereira \etal~\cite{pereira2021procedural_enemies} introducing quality diversity methods, to improve the diversity of the enemies generated. As Pereira \etal, the results show how the generated content matched the desired difficulty.

In 2021, Blasco \etal~\cite{blasco2021evolutionary} looked for generating spaceship enemies which quality is comparable to manually content created by developers. The approach has a novelty as they worked with software models, instead of code. Model-Driven Engineering has the ability to provide an abstraction level in the development process. The results show how the approach was capable to generate content comparable to the manually created by developers in 5 hours, compared to ten months that took to the developers. On other hand, to generate also spaceships, Gallota \etal~\cite{gallotta2022evolving} used a combination of Lindenmayer systems~\cite{lindenmayer1968mathematical} and evolutionary algorithm. Their results suggest that the approach generated spaceships that meet some human preferences.

Some work in the literature tackle battle formations on games where the player must defeat a coordinated group of NPC enemies. Players do not find challenge on the static strategy behind battle formations, as they just learn the optimal way to win. Thus, Ruela \etal~\cite{ruela2012evolving,ruela2014coevolutionary,ruela2017procedural} presented a co-evolutionary approach for generating balanced and challenging battle formations. They evaluated the effectiveness of their proposal offline (i.e., outside the game itself) by comparing the search-based generated battle formations against the battle formations built by human players. The results showed how the generated battle formations could win against the formations of active players, improving the challenge for the players.

A novel work in the literature by Brown \etal~\cite{brown2020evolutionary} proposed a generative approach towards city discovery in four different role playing video games based on the social structures and networks of the NPCs. As a result, the designed algorithm devised the placement of cities and NPCs based on the intern complex relationships of NPCs in order to generate a more realistic video game environment. The evaluation showed the approach to be human competitive.

\begin{mdframed} 
   Most of the work on this type of content focus on enemies NPC, and only one of them on social aspects of NPCs, which provides space to explore more types of NPCs. On other hand, an interesting strategy has been presented by the use of Model-Driven Engineering to generate new content, that could be likely extended to more content types in future work. 
\end{mdframed}
\section{Game Scenarios}
\label{sec:game_scenarios}

Game scenarios describe the goals of a game, and the way and order in which game events unfold. Normally, the events of a game are set in motion as a result of partial or total completion of the game goals, as well as through the interaction between the players and the game. Game scenarios are described by game developers, and are often transparent to the players of the game. Game scenarios can be classified into three different types of content: puzzles, tracks, levels, and stories. Puzzles are problems to which the player must find a solution. The solution can be based on previous knowledge or on a systematic exploration of the space of the possible solutions embedded in the problem. Puzzles can be found in a wide variety of game genres. Some different kinds of puzzles are mazes and physics. Mazes are puzzles defined as a network of paths and hedges through which the player has to find a way. Physics puzzles introduce the laws of the physics into games. Players need to apply the laws of physics for the sake of solving the problem.
Tracks are cyclical game scenarios, and are usually found in racing games.
Levels are logical separators that enable advancement within the different sequences of a game. The advancement is often based on the successful completion of game objectives by the player. Levels can take up many different forms as follows: Rooms are levels where the players must interact with a set of game elements available within a particular section of a game; Dungeons are levels composed by a succession of rooms. The player must pass through the different rooms to complete the goals of the level, Timelines are levels that are linearly designed and are usually found in platform games. 
Stories are the narrative elements that compound the game. Stories present the events of the game to the player affecting directly their experience.
The following paragraphs discuss the relevant search-based research work in each area.

\subsection{Puzzles}
\label{subsec:puzzles}

\subsubsection{Mazes}

A previous work on maze generation by Ashlock \etal~\cite{ashlock2010automatic} presented an evolutionary approach to generate mazes. In the approach, the authors used a direct encoding and a simulation-based and direct theory-driven objective function. In 2011, the same authors~\cite{ashlock2011search} proposed new types of encodings (direct, chromatic, indirect positive, and indirect negative) and features that could be used to construct the objective function. Their results showed that the different encodings and objective functions were feasible, and that the selection of one of them would depend on the desired type of maze. 

In the same year, Ashlock \etal~\cite{ashlock2011simultaneous} presented a work that can generate mazes with two possible solutions, depending on the character that addresses the challenge, named as dual mazes. In this approach, the authors used features for the objective functions from previous work~\cite{ashlock2011search}, but modified the chromatic and indirect positive encodings. The results suggested that the direct encoding generated more diverse mazes and that the indirect encoding found better solutions according to the objective function. Based on this work, McGuinness \etal~\cite{mcguinness2011decomposing,mcguinness2011incorporating} generated large mazes using small mazes as tiles. The novelty of these works~\cite{mcguinness2011decomposing,mcguinness2011incorporating} resided in the objective function, which was adapted in order to provide the developers with more control over the tiles and the final maze. McGuinness \etal evaluated the results of their work by using their approaches to generate large mazes according to designs provided by developers.

In 2012, McGuiness~\cite{mcguinness2012statistical} ran an experiment to statistically compare different encodings from previous work~\cite{mcguinness2011incorporating}. The author argued that the encoding is an important factor to the final visual representation of the mazes. The results suggest that the visual representation of the mazes is very different depending on the encoding, even when the mazes are similar according to the measurements that the author used in the analysis. More recently, McGuinness~\cite{mcguinness2016multiple} built up on previous work by incorporating a direct encoding and features from previous studies in the objective function. In this work, McGuinness~\cite{mcguinness2016multiple} tackled maze generation with a novel search-based approach adapted from the Monte Carlo Tree Search technique. The results revealed that the mazes generated through this approach were intuitive and qualitatively different from those generate by using only evolutionary computation.

Approaches other than evolutionary computation or Monte Carlo Tree Search have also been applied in the last decade. First, Kim \etal~\cite{kim2015quest,kim2018intelligent} proposed a search-based approach to generate `perfect' mazes, that is, mazes with no loops or inaccessible areas. The approach takes as input the desired metrics for the maze and selects the algorithm that better suit the metrics in order to generate the mazes. Once the mazes are generated they are evaluated by the desired metrics and by a set of measure metrics. The difference between the values obtained by the mazes for the desired metrics and for the measure metrics act as the objective function and as the stop criteria for the search. Secondly, Pech \etal~\cite{pech2015evolving,pech2016game} proposed an approach based on the evolution of Cellular Automata (CA) rules that would be in charge of the generation of mazes. The authors argued that evolving CA rules is faster than evolving mazes, because their proposed CA was able to generate a variety of mazes that met the set of rules evolved.

\begin{mdframed} 
  The work on mazes appeared in the literature since the last survey has focused their effort on the representation of the maze problem. This work point out how different types of representation are feasible to generate mazes, but depending on the encoding the results will vary, and while the generated content may be similar based on measurements it can still exhibit visual differences. One open challenge within maze generation is the validation with humans in commercial video games.
\end{mdframed}

\subsubsection{Physics}

Physics-based puzzles are a type of content that has not been tackled by SBPCG before 2013. Shaker \etal~\cite{shaker2013automatic} are the pioneers in this area. They proposed an evolutionary algorithm based on a indirect encoding and a direct objective function. Their results suggested that this technique generated promising puzzles to be played. Afterwards, Ferreira \etal~\cite{ferreira2014generating,ferreira2014search} also presented an evolutionary algorithm for generating physics-based puzzles. However, the encoding and objective function by Ferreira \etal~\cite{ferreira2014generating,ferreira2014search} differ from those proposed by Shaker \etal~\cite{shaker2013automatic}. Ferreira \etal~\cite{ferreira2014generating,ferreira2014search} used a direct encoding and simulation in the objective function, with the main purpose of stability. The results showed that their approach generated stable puzzles.

Kaidan \etal~\cite{kaidan2015procedural} extended these prior work, modifying the objective function to adjust the levels according to the player. Kaidan \etal did a preliminary validation with human players. In a subsequent work, Ferreira \etal~\cite{ferreira2017tanager} built upon their previous work in the field~\cite{ferreira2014search}. In contrast to the work by Kaidan \etal~\cite{kaidan2015procedural}, their approach improved the encoding to allow for more features, such as duplicated blocks. Moreover, the objective function for their approach also took into account the feasibility of a puzzle, not only its stability. The results showed that the approach generated relevant puzzles for the first episode of the game used as case study (\ie Angry Birds).

The direct encoding used in previous work limited the structure of the puzzles~\cite{kaidan2015procedural}. Due to this limitation, Calle \etal~\cite{calle2019free} proposed a novel evolutionary approach to generate stable free form puzzles. In order to reduce the cost to evaluate the objective function, before the simulation, a candidate needed to meet two criteria (distance to the ground and overlapping of blocks). Their results highlighted that SBPCG had potential for generating physics-based puzzles, and the need for problem-specific knowledge. Lately, the same authors~\cite{calle2019speeding,calle2019improved} reduced the cost to evaluate the simulations of the objective function through the usage of a physics engine instead of a game engine.

In contrast with the work mentioned above, other authors~\cite{lara2016spatially,lopez2020checking} tackled the n-body physics problem. To that extent, they proposed an evolutionary approach which generated puzzles according to a given difficulty (easy, medium or hard). The approach by Lara \etal~\cite{lara2016spatially} presented three different objective functions, based on different criteria (intersections, gravitational acceleration, and simulations), and a preliminary analysis was run with human players. The results suggested that none of the three objective functions rated the difficulty of the generated content in the same manner as human players did. Lopez-Rodriguez \etal~\cite{lopez2020checking} further validate the work by Lara \etal with human players. They found out that the automated approach tended to rate the generated content as higher difficulty when compared to the difficulty ratings provided by the human players.

\begin{mdframed} 
  Physics is a novel type of content that we have identified in our survey. It has gained enough attention and several authors have investigated the use of different encodings and objective functions. Some authors recommend the use of indirect encoding to avoid structure limitations on the puzzles. A work with human validation noticed how human and the approach differed over the difficulty of the generated physics puzzles.  Further investigating the reason behind this difference could lead to interesting insights.
\end{mdframed}

\subsection{Tracks}
\label{subsec:tracks}

Some work in the field of tracks~\cite{togelius2006making,togelius2007towards}, included in the previous survey~\cite{Togelius2011}, tackled track generation through an evolutionary algorithm that generated racing tracks for an academic 2D racing game. In order to generate the racing tracks, the authors based their approaches on a bot-simulation objective function to guide the search.

More recent works, such as the one presented by Loiacono \etal~\cite{loiacono2011automatic}, aimed to optimize the fun value of the game through the maximization of the potential diversity of race tracks in the game, namely, through the maximization of the differences between the available race tracks in the game. The diversity of a particular race track was assessed by a multi-objective function that measured the curvature of the track through a direct objective, and the speed profile of the track through a simulation objective. Loiacono \etal performed a preliminary validation with humans, which suggested that there is a statistically significant alignment between the results provided by the approach presented and the preferences of human players.

Prasetya \etal~\cite{prasetya2016search} also worked towards the optimization of the fun of a game through its tracks. They differed from the work of Loiacono \etal mainly in two aspects. Firstly, they used a semi-direct encoding instead of an indirect encoding. Secondly, they compared the performance of two different search algorithms (Tabu Search versus a Genetic Algorithm). Among the two search algorithms, the Genetic Algorithm had less average generation time than Tabu Search. Prasetya \etal performed an evaluation with humans comparing the tracks generated by their approach against man-made tracks. Their results suggested that the generated tracks were measured up to man-made tracks in terms of the measured fun.

With the aim of aligning content generation results with the preferences of the human users of a game, Cardamone \etal~\cite{cardamone2011interactive,cardamone2015trackgen} were the first to introduce an interactive objective function in the track generation process. Through their approach, a population of tracks was generated by an Evolutionary Algorithm. The Evolutionary Algorithm was guided through the assessment of the tracks, provided by human participants after each iteration of the algorithm. After the experiment, humans stated improvements in the quality of the tracks, and that the process produced interesting tracks. 

\begin{mdframed} 
  The work in tracks share a common goal, that is "fun". In order to achieve this objective, human intervention is present either in the evaluation or in the objective function.
\end{mdframed}

\subsection{Levels}
\label{subsec:levels}

\subsubsection{Rooms}
\label{subsubsec:rooms}

We have noticed the need of this subcategory because in the last ten years, the amount of works that fall into this type of content ,and the diversity of techniques (such as Evolutionary Algorithms~\cite{kraner2021procedural}, Multi-objective Algorithms~\cite{deng2021roads}, Quality Diversity~\cite{green2022persona}), has increased. As an example, in 2012, Togelius \etal~\cite{togelius2012compositional} proposed a preliminary approach that tackled this type of content through a hybrid approach that used Evolutionary Algorithms and Answer Set Programming (ASP), which has been used before for system design generation~\cite{smith2010variations}. However, in more recent work, the core of the generation of this type of content has shifted towards the usage of the General Video Game AI (GVGAI) framework. Two main elements are used to build this framework: (1) the Video Game Descriptive Language, and (2) the General Video Game Playing Competition. The Video Game Descriptive Language (VGDL)~\cite{schaul2013video} is a textual description language that has been used to represent two-dimensional games. The General Video Game Playing Competition~\cite{perez20152014}, which started in 2014, is an event that explores the challenge of creating controllers for general video game play, where a single agent must be able to play many different games. The GVGAI framework provides a series of different games based on VGDL, as well as the game-independent agents to play the generated room levels for those games.

Neufeld \etal~\cite{neufeld2015procedural} used an Evolutionary Algorithm to evolve the rules used by a room generator from GVGAI based on ASP. These rooms were evaluated through a simulation objective function that calculated the difference of average scores obtained by vanilla Monte Carlo Tree Search and a random player. Their results showcased the benefits of the approach, however, Drageset \etal~\cite{drageset2019optimising} identified the computational cost of translating VGDL games into ASP rules as a drawback. Hence, Drageset \etal proposed a purely evolutionary approach, named Meta Generator, based on a more elaborated simulation-based objective function that uses three of the agents provided by GVGAI. The evaluation compared the Meta Generator against both random and constructive generators using the same objective function. The results highlighted that the Meta Generator obtained higher scores for the objective function than the other generators. 

Zafar \etal~\cite{zafar2020search} also proposed an evolutionary approach, more precisely, a Feasible Infeasible Two Population algorithm which differ from previous work in the metrics used for the objective function, measuring aesthetics and the difficulty of the rooms. Their results suggested that the levels obtained were aesthetical and challenging. With the same aim, Petrovas \etal~\cite{petrovas2022procedural,petrovas2022generation} proposed a genetic approach, with a complex direct fitness function, Combined Compromise Solution.

Another proposal from Zafar \etal~\cite{zafar2019using} used design patterns to generate rooms. The approach selected patterns from a collection of design patterns and used them as input for the evolutionary process. The objective function also included a factor in the equation related to design patterns. The experiments were run with different agents from GVGAI, and concluded that the agents had better performance on rooms generated with design patterns.

Walton \etal~\cite{walton2021evaluating} addressed the room generation from the perspective of the developers instead of the player. Their proposed approach takes as input a level designed by the developer, and uses a Feasible Infeasible Two Population algorithm to generate new levels. This approach was evaluated by developers judging its capacity to facilitate their job. The developers found more inspiring the results from the evolutionary algorithm than random generation. However, the approach lacks of diversity, which limits its use.

In order to tackle this issue, which is common to different approaches, a novel evolutionary strategy, named Illumination Algorithm\cite{mouret2015illuminating}, was exploited. Illumination Algorithms find high performing solutions in different sections of the search space instead of maximizing one solution as evolutionary algorithms usually do. Charity \etal~\cite{charity2020mech} incorporated this idea in their approach, and used the mechanics from some of the games provided by the GVGAI framework to create multiple, relatively high quality states for a GVG-AI level that demonstrate combinatorial variations of a game's mechanics.
Their results showed that this approach generated satisfactory rooms with a single mechanic or with a controlled combination of mechanics.

The use of Quality Diversity methods has grown also in the field of room generations. Green \etal~\cite{green2022persona} introduced Constrained MAP-Elites, generating rooms based on different human play-styles called `Personas'. Using Personas to generate content can encourage a player towards new challenges. Alvarez \etal~\cite{alvarez2020interactive} and Charity \etal~\cite{charity2020baba} proposed a co-creative approach, where MAP-Elites approach proposes new rooms and users guide the generative process. Users can also modify the generated content influencing the evolutionary process.

In contrast with the work study above, Bhaumik \etal~\cite{bhaumik2020tree} were curious about the performance of more Search-Based strategies. Due to that reason, Bhaumik \etal~\cite{bhaumik2020tree} compared eight different search-based algorithms, including Tree Search Algorithms and Optimization Algortihms. Those algorithms inlcude Breadth First Search, Depth First Search, Greedy Best First Search, Monte Carlo Tree Search, Hill Climbing, Simulated Annealing, Evolution Strategy, and Genetic Algorithm. Their results suggest that Optimization Algorithms generally performs faster than Tree Search Algorithms.

\begin{mdframed} 
  Most of the work in this category is related to the General Video Game AI, and its interest on the General Video Game Playing Competition. However, after a human-based evaluation unveiled  that the generated content lacks of diversity, researchers seems to have moved towards addressing this issue. This also suggests that a human evaluation is strongly recommended for all the approaches based on diversity.
 \end{mdframed}

\subsubsection{Dungeons}

Dungeons play an important role in video games. In the last ten years, the interest has not decreased. One of the challenges of generating dungeons is the reduction of the generation time. Pereira \etal~\cite{pereira2018evolving} claimed that a tree structure reduce the need for validation and fixing time that graph/grid approaches required, such as the one proposed by Valtchanov \etal~\cite{valtchanov2012evolving}. Later, Pereira \etal~\cite{pereira2021procedural} conducted an experiment with human players to validate the generated dungeons. The results showed that the dungeons were enjoyable and challenging.

Font \etal~\cite{font2016constrained} used a graph approach to reduce the generation time, which differs from the work by Valtchanov \etal~\cite{valtchanov2012evolving} for two aspects. First, the approach used a context-free grammar representation to avoid the generation of syntactically non-valid individuals. Secondly, Font \etal~\cite{font2016constrained} reduced the search space by dividing the approach in two steps: Generating the structure of the dungeon first, and the detailed elements that are more time consuming, such as monsters or chests, afterwards.

Another challenge in the generation of dungeons is the diversity of the results. Tackling this challenge, Ruela \etal~\cite{ruela2018scale,ruela2018evolving,ruela2020multi} proposed a single-objective approach, which later derived into a multi-objective approach due to the fact that a single-objective approach that combines different objectives tends to prioritize one in detriment of the others. Melotti \etal~\cite{melotti2018evolving} proposed a variation of a multi-objective approach combined with Deluged Novelty Search Local Competition, which separated the search space into niches, allowing for the control of the differentiating characteristics of the niches through a distance function.

On a different perspective, Liapis~\cite{liapis2017multi} addressed dungeon generation using intertwining segments as representation. And, inspired by Liapis representation, Viana \etal~\cite{viana2022feasible} generate dungeons using barriers as novelty, which are mechanics that force players to follow a path. The diversity of the dungeons are measured by map linearity, mission linearity, leniency, and path redundancy. Their results analysed through expressive range suggest that the approach achieve the diversity on dungeons generation.

In addition to diversity, Ruela \etal~\cite{ruela2020multi} designed an experiment to compare the performance of four well-known algorithms against their own algorithm. They concluded that it is not possible to define an overall winner algorithm, and that the use of each algorithm will depend on designer preferences. The baseline algorithms were: Improved Strength Pareto Evolutionary Algorithm (SPEA2), Pareto Archived Evolution Strategy (PAES), Non-dominated Sorting Genetic Algorithm II (NSGA-II), and Multi-objective Cellular (MOCell). The main limitation of the novel algorithm proposed by the authors was its time consumption, which limited its usage to offline PCG.

To scale up the resources of a game, Brown \etal~\cite{brown2017procedural,brown2018levels} proposed the integration of a dungeon editor that players can use into a commercial game. The proposed search-based approach generated different rooms within a space and then connected the rooms. The connection was possible due to the hard constraint that each new room must overlap with an existing one. This work serves as an indicative that hard constraints are an advantage that search-based methods provide as we mention in the ML constraints (Section~\ref{sec:introduction}). 
Brown \etal also allowed the integration of objects and enemies into the dungeon. A Petri net method filled the dungeon with objects, and a second evolutionary approach placed enemies throughout the dungeon.
On other hand, Harisa and Tai~\cite{bagus2022pacing} generate dungeon levels for based on game designer preferences of game pacing. The experiments showed the error between the designer preferences and the generated content with results between 1.16-18.53\%.

\begin{mdframed} 
  The work on dungeon generation has identified two challenges for this type of content. First the need to reduce the generation time, and secondly the diversity of results. Due to those challenges the generation of dungeons has mainly remained an offline generation. However, there is work on online generation, which makes use of hard constraints in the approach. There is no outcome from online approaches about generation time or diversity.
 \end{mdframed}

\subsubsection{Timeline}

Most of the work under this category used a well known platformer video game as a case study, Infinite Mario Bros, which is a clone of Super Mario Bros. However, we have decided to name this subcategory as "Timeline levels" because a more general name provides the potential of covering a wide range of genres, such as platform games, runner games, or endless games. Shaker \etal~\cite{shaker2015progressive} is a good example of a generic approach of this category. The approach generated linearly the sequence of actions for different games. They validated the approach simulating the sequences.

An advantage of the timeline levels is that they reflect the difficulty curve of games. A difficulty curve is a graphical representation of how the difficulty fluctuates during the game. Several approaches has addressed this challenge on platformer games~\cite{adrian2013approach,moghadam2017genetic,kholimi2018automatic}. Adrian \etal~\cite{adrian2013approach} was the first one that used the difficulty curve designed by developers as objective function. The experiments showed that the difficulty curve of the generated timeline levels were close to the designed difficulty curve. A preliminary validation with 22 players supported the similarity with handmade timeline levels. Inspired by Adrian \etal, Atmaja \etal~\cite{atmaja2020difficulty} applied the same idea into a scrolled vertically shooter game. Moghada \etal~\cite{moghadam2017genetic} also generate the rhythm of the level, fitness with difficulty curve by human. 

Different representation approaches has been studied for timeline levels, more precisely for Mario Bros. In 2011, Mourato \etal~\cite{mourato2011automatic} found out that the use of a grid as a detailed representation of timeline levels had the risk of consuming substantial  resources. To reduce the consumption of resources, Dahlskog \etal~\cite{dahlskog2013patterns,dahlskog2014procedural} divided Mario Bros levels into vertical slices. Each slide was a micro-patterns that they extracted from the original video game. The combination of the slides generated the resultant timeline levels, named as "scenes". Dahlskog\etal~\cite{dahlskog2014procedural}  validated the approach finding in the generated scenes combinations of micro-patterns that expressed meso-patterns that were originally in the game.
Later, Green \etal~\cite{green2020mario} "stitched" the scenes generating long timeline levels.
In 2022, Moradi \etal~\cite{moradi2022using} introduced Estimation Distribution Algorithm to also create meso-patterns.

Based on the work of Dahlskog \etal, Green \etal~\cite{green2018generating} generated scenes that served as tutorials for a specific mechanic. The approach used a feasible infeasible two population algorithm with two objectives function. The infeasible population objective function measured the aesthetics, e.g. a pipe on Mario Bros requires two consecutive slides. The feasible objective function compared the performance of two agents, a "perfect" agent and a limitated agent. When the limitated agent failed on completing the level, it meant that a special mechanic is needed, and therefore the level could be a tutorial.
Khalifa \etal~\cite{khalifa2019intentional} compared three approaches which generate timeline levels for one mechanic, including the work of Green \etal~\cite{green2018generating}. They results showed that the approach that guarantee the mechanic on the scene had three disadvantages: It was the slowest approach, required human intervention, and relied on agents failures.

We have seen that in more recent years, new techniques such as illumination techniques (see Section~\ref{subsubsec:rooms}) have appeared. Warriar \etal~\cite{warriar2019playmapper} presented a first attempt of Illuminated approach for platform video games. The results highlighted three opportunities for improvement: A more visual human-designed aesthetic levels, better features that define fun, diversity and controllability on the content, and how to keep a computationally low consumption with an entire map of playable levels.
In this direction, Withington~\cite{withington2020illuminating} presented a preliminary comparison of Quality Diversity algorithms MAP-Elites~\cite{mouret2015illuminating} and SHINE~\cite{smith2016rapid}, however neither approach stood out above the other.

\begin{mdframed} 
 Research on timeline generation has worked towards three different aspects. The first one is the use of a difficulty curve designed by developers to approximate the generated content to the desire curve. Second is the effect of using different representations. The last one is seeking diversity over the results. These three aspects has been addressed separately, where future work it is needed to investigate their combination.  
\end{mdframed}

\subsection{Stories}
\label{subsec:stories}

Addressing the generation of stories is a complex task for procedural content generation. The reason is the difficulty to obtain a cohesive story from evolutionary operators. However, the use of trees or grammars with small pieces of stories are known to work well together~\cite{mason2019lume}. Based on the use of trees and the recent works to reach diversity through Search-Based PCG, Fredericks \etal~\cite{fredericks2021genetically} introduced a novel idea that combines those two elements into a genetic improvement algorithm. The work is still under development, and currently lacks of evaluation.
In this direction, Alvarez \etal~\cite{alvarez2021questgram} and De Lima \etal~\cite{de2022procedural} has also generated narrative quests making use of grammars~\cite{alvarez2021questgram}, and trees~\cite{de2022procedural}.

By the hand of Alvarez \etal~\cite{alvarez2022tropetwist,alvarez2022story} Quality diversity methods has also been applied in stories generation. Alvarez \etal~\cite{alvarez2020interactive} first proposed a co-creative approach to generate rooms (See \ref{subsubsec:rooms}) that has been adapted and applied for stories~\cite{alvarez2022story}.

\begin{mdframed} 
  Even with the difficulties that stories present to PCG, novel ideas have been applied for this type of content. Most of the work move toward achieving diversity in the results.
\end{mdframed}
\section{Game Design}
\label{sec:game_design}

Game design is the core of a game. Game design defines the gameplay of a game by conceiving and designing its rules and structure. A change in game design could generate a whole new game. Game design decisions affect all the content in the previous sections. We can further distinguish game design into two aspects: (1) System design which refers to the rules and mechanics that define a game; (2) Camera control which refers to the placement and behaviour of the camera in the game, or in other words, how the player will visualize and experience the game.

\subsection{System Design}
\label{subsec:system_design}

As seen in the previous SBPCG survey~\cite{Togelius2011}, system design has been addressed by different approaches. 

Browne~\cite{browne2008automatic} presented the Ludi system, which used Evolutionary Computation combined with a simulation-based objective function, with the aim of measuring aesthetic aspects of the game. This system led to the first fully computer-invented games to be commercially published. Togelius \etal~\cite{togelius2008experiment} introduced the concept of `fun' measurement through a Hill-Climbing approach. Both approaches used simulation-based objective functions that have been used in recent works to tackle system design from different points of view.

Aligned with the work by Togelius \etal, Halim \etal~\cite{halim2014evolutionary} guided the search with the aim of optimizing entertainment. The objective function measurements included the duration of game, the intelligence required to play the game, the dynamism exhibited by the pieces, and the usability of the play area. Halim \etal validated their approach by conducting two experiments: one with a neural-network-based AI to measure controller learning ability, and another with human users to measure the entertainment provided by the game. The results suggested that the evolved games were more interesting and better than randomly generated games.

Kowalski \etal~\cite{kowalski2015procedural,kowalski2016evolving} proposed different approaches to generate games based on chess. In 2015, Kowalski \etal~\cite{kowalski2015procedural} presented an objective function based on simulation to generate playable and balanced games. A hand-made evaluation function analysed playout histories and checked the balance, game tree size, pieces importance, and complexity of the game rules. In 2016, Kowalski \etal~\cite{kowalski2016evolving} generated games that were closer to chess and more constrained than those generated by their previous work. The objective function increased the number of agents in the simulation, representing players with various degrees of intelligence. The experiment had two parts, the first compared the different agents of the simulation, and the second compared the results with human-made chess-like games. The results showed that the approach obtained playable and balanced games that were similar to the high quality human-made games.

\begin{mdframed} 
    In  recent years, three different goals have been addressed by System Design approaches. First, tackling aesthetics, secondly, optimizing entertainment, and last, playability and balance. 
    The latest search strategies rely on Quality Diversity, which is a promising opportunity fo generating this type of content.
\end{mdframed}

\subsection{Camera Control}

Previous work on automatic generation of camera control content~\cite{burelli2010combining,burelli2010global} showed that the search space of this type of content is rough to be explored. In addition, it has also been highlighted that the objective function for generating camera control content is computationally expensive, which reduces the number of evaluations available for the search process. 

Preuss \etal~\cite{preuss2012diversified} addressed automatic camera control through a niching and restart Evolutionary Algorithm. Niching extends Evolutionary Algorithms to multi-modal domains, locating multiple optimum candidates where the Evolutionary Algorithm loses population diversity, converging to a unique solution. Convergence is an issue that Preuss \etal tackled through a restart mechanism when the approach reached this point. Preuss \etal improved their previous approach~\cite{preuss2010niching} by adding a constraint that reduced the search space, which is one of the problems in automatic camera control content generation. Their evaluation compared their approach and state-of-the-art algorithms. Their results suggested that their novel approach and another similar approach (CMA-ES) performed better than other prior approaches.

\begin{mdframed} 
    Camera control novel approaches are performing better than prior approaches, at least in academic settings, which makes it interesting to further asses the performance for commercial games.
\end{mdframed}
\section{Future Directions}
\label{sec:fd}

Throughout the pages of this section, we analyse the different research challenges that are open in the field of work of PCG for video games. To that extent, the following subsections present the challenges, their reason to be, and the open problems for each challenge, as well as recommendations and potential future research lines in the identified areas.

\subsection{Content Opportunities}
\label{sec:fd:co}

\begin{figure}   
    \centering
    \caption{Number of articles published per year for each category of the taxonomy studied in this survey.}
    \includegraphics[width=0.4\linewidth]{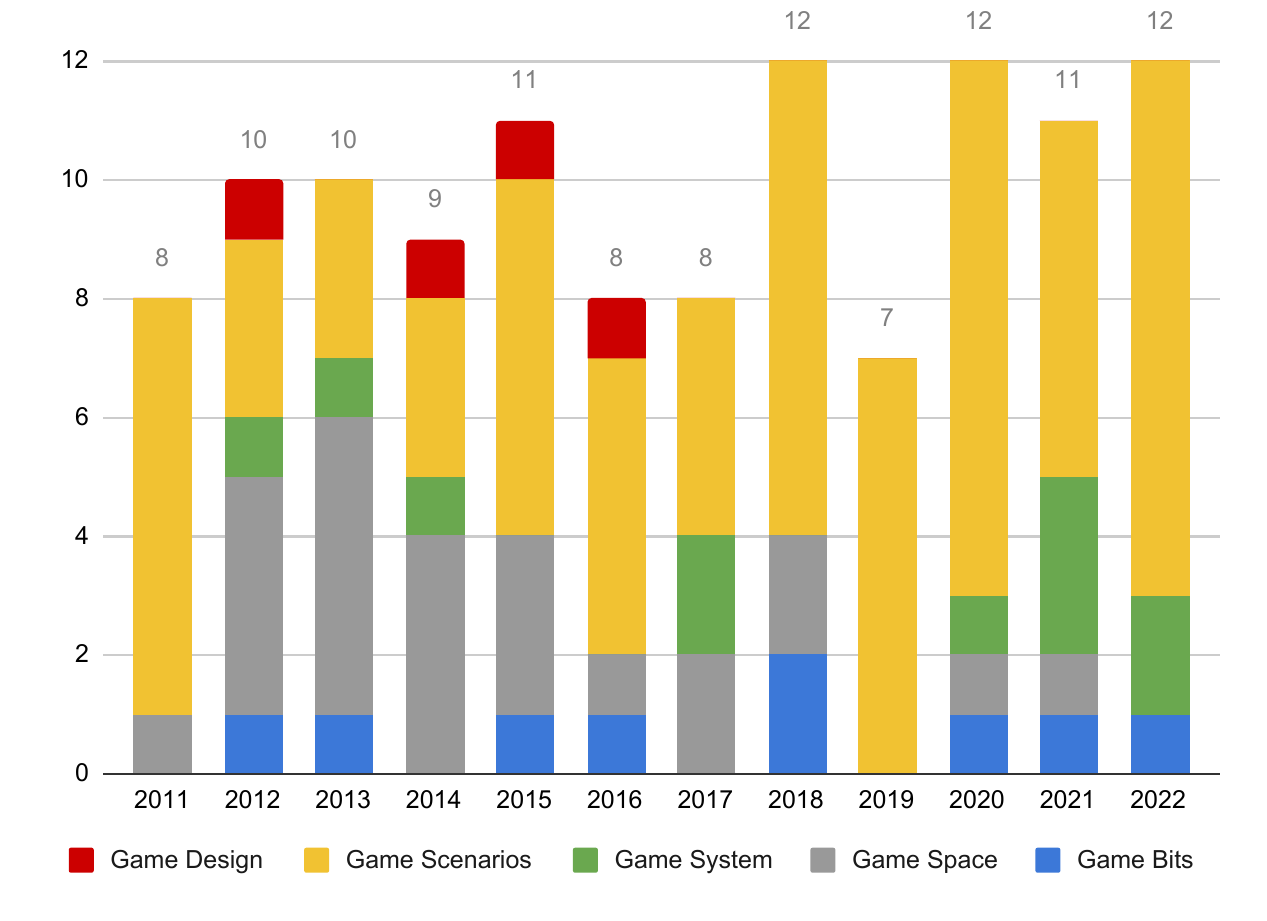}
    \label{fig:wpy}
\end{figure}

Our survey has shown that the number of publications is not balanced across the content types. In Figure~\ref{fig:wpy}, we show the number of articles published per year that tackle the generation of content for each category analysed in our survey. While Game Scenarios and Game Spaces arise as the most popular content types, it is also possible to identify three exceptionally unpopular content types: Game Bits, Game Systems, and Game Design. Regarding the Game Bits category (Section \ref{sec:game_bits}), which comprises the most basic building blocks of a game, we find a total of 7 works. Within the category, we find the textures (4 works), sound (a single work), and weapons (3 works) subcategories. Regarding the Game Systems category (Section \ref{sec:game_systems}), which deals with bringing virtual worlds closer to the human world through the study of NPCs, we find a total of 11 works. Finally, regarding the Game Design category (Section \ref{sec:game_design}), which defines the core of a game in the form of its rules and structures, we find again a total of 4 works dealing with system design (3 works) and camera control (a single work). The study of sounds and camera control are, then, the most neglected content types among already unpopular categories. However, we could not find a clear reasoning coming from the community for the lack of work in those particular areas, so we can only speculate why they are being neglected. Games and video game genres appearing in a recent report on the status of the video game industry\footnote{\url{https://www.wepc.com/news/video-game-statistics/}} include those types of content, so their necessity in video game development becomes apparent. As a matter of fact, we have considered that some of these content types have been neglected because they are considered too important and integral to the success of a video game: interactions with NPCs, for instance, help players with video game immersion; and poor sound effects can lead to a worsening of the user experience, which can in turn lead to poor reviews and poor sales as a ultimate consequence. This may be a reason that calls for a special treatment, and for a careful development by hand. We also theorize that there may be too much diversity in the possible outcomes of PCG in those areas, which makes it easier for researchers to focus on other content types.  

In any case, the existence of fewer studies in those areas implies that there are less improvements on the current complexity of their generation process and on the complexity of the needed constraints. This issue could in turn mislead the directions of researchers that favour other content types, since the current state of research makes it possible to believe that Search-Based approaches are not suitable for these content types even when nothing appears to indicate that those areas are harder to generate through SBPCG than other content types. In that sense, we believe that these content types should not go unnoticed by the research community any longer. Even if there is an inherent diversity to the content, many SBPCG approaches allow constraints to the objective function or different degrees of customization that may be useful to tailor the results to the necessities of each case study. In addition, as already discussed, details as textures, sounds, or interactions with NPCs are essential to captivate players and provide quality to a video game. A game that lacks these content types is directly headed to a failure. Hence, the procedural generation of these content types could not only help with the economic cost and time to market of a video game, but also with the outright feasibility of a game where developers lack the means to build some of these content types. 

Finally, regarding the suitability of SBPCG for these content types, we want to encourage researchers to avoid steering their research to those fields of work that are already explored, and raise awareness about the opportunities of research in the field. In particular, we recommend to revisit approaches that have proven their potential towards PCG applications, and to explore their application towards researching the neglected content types. Overall, we recommend to approach all content types equally, as they are equally important for the development of a video game.

\subsection{Online PCG}
\label{sec:fd:opcg}

The content of a video game can be added at two different points of the development process. Those two points are either before the release of the game, offline generation, or during play, online generation. Most of the works presented in this survey work with offline generation approaches. We speculate the reason behind this lies in the content generation time issue: the current reality is that approaches take too much time to generate content to be viable in online contexts. This problem affects offline generation to a lesser extent because there is more flexibility with the time for content generation during the development process. This issue will stand for online content generation as long as approaches lack the necessary speed to not paralyze the experience of the player. 

The content generation time issue finds its roots mainly in the inefficiency of the approaches and the computational resources that those approaches use. As an example, approaches that use an indirect representation of the content require less resources, however, indirect representations require a transformation process that turns the indirect representation used in the approaches into the final content that is incorporated in the video game. On the other hand, simulation objective functions require more time than direct objective functions. Games that actually use online generation do so through the combination of preexisting elements, with the disadvantage that the content is then limited to designs that have already been established by the developers offline. 

The challenge lies in the application of current approaches to online generation. To that extent, it would be necessary to identify applicability problems of current approaches, improve the necessary generation times, and optimize the device resources used by a game to dedicate more computing time to the generation of content. In addition, to study these issues with a greater level of detail, it becomes necessary to know the time that takes to generate the content. However, most of the presented works in our survey do not provide the times associated to the generation time. We recommend to report the generation time in the results of future work in this field. 

Finally, we believe that it would be possible to reduce or even avoid these challenges through the usage of remote servers. Ideally, the remote servers would run the approaches while the players play the game, generating the content in parallel. In theory, this would allow for online generation while the players make full usage of their playing time. Exploring the usage of remote server SBPCG for online content generation has been largely neglected in research in the field, save for works in tracks~\cite{cardamone2011interactive} that explores remote generation of content, and hence, it remains a promising direction for solving the challenge.

\subsection{Solvability, Playability, Fairness, and Diversity}
\label{sec:fd:spfd}
Through the literature in the field we identify solvability, playability, fairness, and diversity as the measurements that identify the basis of player expectations towards a game. Solvability is understood as the characteristic that defines whether a problem presented in the game content can be completed or solved (e.g. going from point A to point B in a level). Playability is understood as the measurement that defines the extent to which content can be exploited by human players. Fairness deals with the perception of the player when dealing with content (e.g. distribution of resources in a strategy game, or the probability of an event occurring in the game). Diversity describes the variety of the content so that players do not receive similar content. All of these factors affect the overall feelings of players towards a game, and influence the decision on whether to keep playing or not. For instance, it is important for a game to be challenging but not impossible to finish, and unfair game mechanisms generate frustration in players.

Due to the reasons listed above, these measurements are commonly used to evaluate the results produced by approaches and to guide them in procedural content generation. However, they are not the only means to evaluate the results of the objective functions in the available research. To that extent, we have observed multiple research directions regarding the objective function in use for the evaluation of the generated contents. In that sense, while some works produce objective function scores based on the above metrics, other works retrieve their evaluation from the behaviour of simulators that intend to substitute human players, and finally, the works that obtain the best results in the literature make direct use of humans as the objective function function in what is known as interactive objective function.

While interactive objective function leads to the results better aligned with the expectations of players, it is still not a perfectly adequate objective function. As a result of experimentation, human participants tend to fatigue, which leads to a worsening of the objective function over time and a ceiling effect in the results that hinders the potential of research in the field. In addition, the fatigue of human participants limits the application of approaches to advanced or complex case studies, which we can observe in the fact that most of the studies in this field evaluate the approaches over severely tailored academic case studies.

As a potential research direction, we recommend revisiting objective functions and their application to the research field. In that sense, we believe that researchers should explore improvements to interactive objective function through the incorporation of mechanisms to avoid fatigue. To that extent, it would be possible to explore hybrid objective functions that combine metrics or simulators (or both) with interactive objective function. The combination would see metrics or simulators working for a while on their own, guiding approaches towards preliminary results that could be assessed afterwards by human hands. Exposing humans to short interactions with an evolutionary approach at suitable times would avoid fatigue, thus allowing the application of the approaches to the more complex industrial case studies, opening promising lines of research in real-world environments.

In addition to the above, in the case of offline generation, if results are not satisfactory, it is possible to generate new content or involve developers to refine the content manually. This is not possible for online content generation. Moreover, the prior survey~\cite{Togelius2011} expressed concerns with the limitation of diversity that may be caused by approaches looking for the best possible results. To overcome this limitation, research in the past few years has seen a surge in quality diversity approaches~\cite{gravina2019procedural}, a young and promising field that has attracted the attention of PCG researchers (on categories such as weapons, strategic maps, timeline or rooms), and is yet to be applied to several content types in both offline and online generation. A hybrid objective function that combines different measurements with human evaluation, along with the possible generation of content in remote servers mentioned as part of another future direction (Section~\ref{sec:fd:opcg}), would help with the evaluation of content generated online, and with the acquisition of feedback directly from the players, who are the best source of information for indicating the viability and diversity of the generated content. 

\subsection{Bricolage}
\label{sec:fd:b}

The main goal of PCG is to help developers during the development process of the content of a game. However, not all generated contents are fit to be directly included in a game. In that sense, the content that is created, stored, but never used creates a waste of resources. In addition, discarded content becomes a potential source of frustration for developers, who may evaluate those contents in terms of working hours, or even see them as promising ideas thrown away, only unfeasible because of the amount of time that it would take to fix them by hand so that the content can make the cut into the game.

However, we could tackle this issue from a new angle, considering the contents discarded after a content generation process as material that can be potentially used to refine and adapt the process to generate suitable content for the game. To that extent, it would be possible to consider content as a sum of components rather than as a whole, and to evaluate the suitability of each component for the game independently. In that way, it would be possible to explore the reuse of components from existent content to novel content, thus avoiding waste of resources and the frustration of the developers. 

The reuse of components would also not be limited to discarded content, on the contrary, it would be possible to take into account components coming from all the existing content for a video game. For instance, reuse  of components could leverage components from content that is under development in order to generate refined versions or variants of the content during the development process, allowing developers to have a wider choice of content. Reuse  of components could also use components from content that has been already approved into a game, gaining access to a library of components that count with the endorsement of the developers of the game. In addition, though the usage of reuse  of components, developers could be more involved with the generation process, choosing their favored components to refine the directions of the approaches. Our recommendation for this research direction is to build approaches, both for online and offline PCG, that take developer involvement into consideration and that empower the reuse of components of the existing content.

\subsection{Statistical Rigor}
\label{sec:fd:sr}

We have identified two common practices towards evaluating the presented approaches. The first of them is the usage of an objective function as a measure to evaluate the outcomes and reliability of the approaches. However, trusting the objective function requires a prior evaluation of the objective function, and causes the reliability of content generation systems to fall on the reliability of the objective function. The second one is to present the approaches to events in order to compare their efficiency. In this case, the results of the approaches are set against each other, limiting the comparisons between approaches to the rules of the event and not specific measurements. In such scenarios, it becomes necessary to improve the execution of the approaches and their comparison. In that sense, it would be advantageous to have a baseline that enables a fair comparison between works and equal opportunities for progress in the community, as well as a fair comparison of certain aspects that could not be compared otherwise. 

In addition to the above, the criteria followed by events as a means of evaluation is not uniform. While some events leverage results from bots to evaluate the approaches, the usage of feedback provided by real players is gaining momentum in the field in the form of interactive objective function. We can find an example of the latter in an event where a jury provides an assessment of the results of an approach based on the performance of players. Many good practices in this field of work, such as A-B testing, could be leveraged to build hybrid interactive systems that help with the involvement of players and developers while avoiding fatigue. Overall, we advocate for building approaches that allow practitioners and developers alike to get closer to the target public, to ensure that the generated content can live up to the expectations of players. 

Finally, through this survey, we found out that many studies refer to artifacts and results not publicly available, which severely hinders research replication. Following the principles of open science\footnote{\url{https://github.com/acmsigsoft/open-science-policies}} to which the general Software Engineering community adheres to, we strongly recommend the publication of prototypes, approaches, artefacts, and research results. This would not only help with research replication, but it would also influence the growth of the community and the rigor of the presented research.

\subsection{Industrial Content}
\label{sec:fd:ic}

\begin{figure}   
    \centering
    \caption{Percentage of articles where the case study involves academic games, academic environments or commercial games.}
    \includegraphics[width=0.4\linewidth]{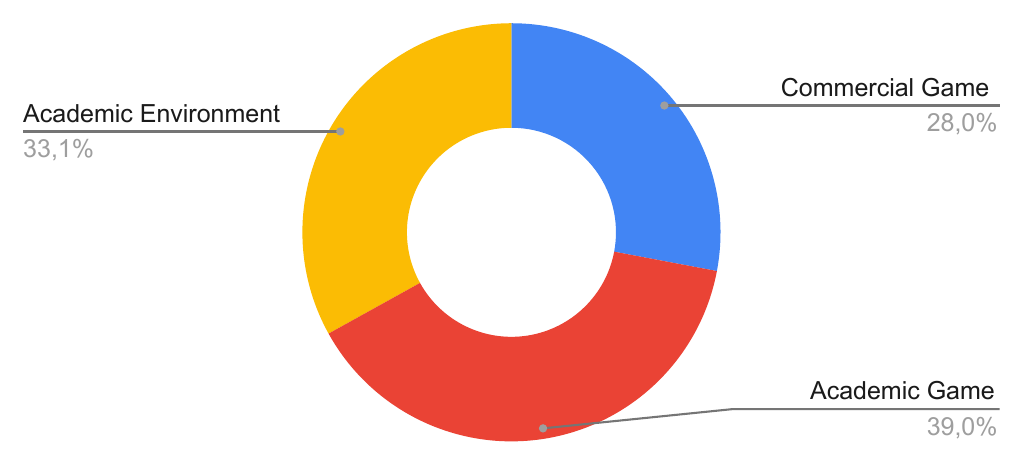}
    \label{fig:games}
\end{figure}

The majority of the studies presented in this survey focuses on academic games or academic environments as case study (see Figure~\ref{fig:games}). Those academical games are clones of industrial games, simplifications or prototype versions of original games. Academic environments are environments built explicitly for the research purposes of the researchers. Plenty of research questions arise from the observed scenario: Is there a need of more detailed generators? Is it difficult to adapt the representations used in research to industrial content? Is the content of industrial games more complex, making the quality of the generated content insufficient in industrial scenarios? Whatever the case, the reality is that research in the field is rarely applied to real-world video games. With such a scenario, the research community remains disconnected from the industry, causing poor communication of results between the novelties in SBPCG research and the final target users of those approaches, the players. In other words, current research results do not reach players. For example, Ruela \etal~\cite{ruela2012evolving,ruela2014coevolutionary,ruela2017procedural} could not empirically assess their proposal involving humans due to the lack of access to the original game developers, and the effort and time that human players would have had to put in for evaluation

This issue leads to what we believe is one of the major opportunities for research in the field. If we are able to avoid this disconnection by applying the approaches to real-world industrial case studies, we might be able to obtain feedback from players, which might represent a very large and very valuable source of information for developers and researchers alike. This source of information could be used to guide automated approaches and to manually refine the generated content. However, in order to apply the approaches to industrial content, we must avoid fatigue. In addition to the exploration of hybrid interactive approaches, in blockbuster games with millions of active players, it would also be possible to research mechanisms to share the fatigue load of the fitness function. To that extent, our recommendation is to identify the necessities of developers working in industrial contexts, and to adapt the proposed approaches so that they can work over commercial content.

\subsection{Interaction between SBPCG and other techniques}

Throughout the survey, we have focused on pure SBPCG approaches that generate content. However, there exist work that tackle PCG through the interaction between SBPCG and other techniques, specifically ML-based techniques. In the recent years the interaction between SBPCG and PCGML has gained interest due to a new research line called latent variable evolution~\cite{volz2018evolving}. Latent Space (LS)~\cite{bontrager2018deepmasterprints} allows to learn the shape of search spaces where later the approach can search more effectively. This is important, for example, for online PCG, where the time of the search matters. On the other end, Quality Diversity (QD), a novel research field in SBPCG, has already gain attention from the PCGML community exploring the interaction between QD and LS~\cite{fontaine2021illuminating}.

\section{Conclusion}
\label{sec:con}

The high demand for video game content has led to an increased interest in PCG, and its investigation has gained momentum in the past decades. 
Throughout the pages of this article, we have surveyed the updates in the state-of-the-art stemming from the 10-year gap since the last surveys in PCG for video games were published. To that extent, we have built a taxonomy based on the two prior surveys~\cite{Togelius2011,Hendrikx2013}, and gathered and categorized novel research in SBPCG. As a result, we have reported  herein on new work in Game Bits, Game Space, Game Systems, Game Scenarios, and Game Design. Despite the undeniable advances in research in the field, we consider that there are still many unexplored topics, and that some of the research challenges and recommendations proposed 10 years ago are yet to be studied in depth. Through our work, we have identified plenty of open research challenges regarding content opportunities, the speed of online PCG, the characteristics (solvability, playability, fairness, and diversity) of the generated content, the possibilities enabled by content bricolage, the inclusion of statistical rigor in this field of research, and the need for applying research to industrial content. Along with the open research challenges, we have presented recommendations and identified several potential future research lines in the areas under study. Overall, this survey presents a concentrated and comphrensive report on the latest work in SBPCG , effectively assessing the status of research in the field and providing a renewed point of view for the ongoing discussion over SBPCG in video games.



\section*{Acknowledgments}

Work supported by National R+D+i Plan  PID2021-128695OB-100,  José Castillejo CAS18/00272,  ERC grant 741278.

\bibliographystyle{ACM-Reference-Format} 
\bibliography{references.bib}

\end{document}